\documentclass[letterpaper,aps,prc,superscriptaddress,showpacs,nofootinbib,floatfix,twocolumn]{revtex4-1}

\usepackage{hyperref}

\usepackage{color}
\usepackage{graphicx}	
\usepackage{chngpage}
\graphicspath{%
  {./},%
  {./figures/}%
}

\usepackage{xspace}	

\newcommand{\pt}{\mbox{$p_{_T}$}\xspace}

\newcommand{\rdau}{\mbox{$R_{d\rm{Au}}$}\xspace}
\newcommand{\rdauave}{\mbox{$R_{d\rm{Au}}$(0-100\%)}\xspace}

\newcommand{\rda}{\mbox{$R_{dA}$}\xspace}
\newcommand{\rpa}{\mbox{$R_{pA}$}\xspace}
\newcommand{\rcp}{\mbox{$R_{\rm CP}$}\xspace}

\newcommand{\dau}{\mbox{{\it d}+{\rm Au}}\xspace}  
\newcommand{\pau}{\mbox{{\it p}+{\rm Au}}\xspace}  
\newcommand{\pA}{\mbox{{\it p}+{\it A}}\xspace}  

\newcommand{\sqsn}{\mbox{$\sqrt{s_{_{NN}}}$}\xspace}
\newcommand{\jpsi}{\texorpdfstring{{\mbox{$J/\psi$}}}{J/psi}\xspace}  

\newcommand{\ccbar}{\mbox{$c\overline{c}$}\xspace}
\newcommand{\sbr}{\mbox{$\sigma_{br}$}\xspace}
\newcommand{\mrt}{\mbox{$M(r_{\scriptscriptstyle T})$}\xspace}
\newcommand{\lrt}{\mbox{$\Lambda(r_{\scriptscriptstyle T})$}\xspace}
\newcommand{\rt}{\mbox{$r_{\scriptscriptstyle T}$}\xspace}
\newcommand{\ttt}{\mbox{$2 \rightarrow 1$}\xspace}

\newcommand{\ntub}{\mbox{$N_{tube}$}\xspace}
\newcommand{\rtub}{\mbox{$R_{tube}$}\xspace}
\newcommand{\x}[1]{\mbox{$x_{#1}$}\xspace}
\newcommand{\qsq}{\mbox{$Q^2$}\xspace}

\begin{document}

\title{Theoretical Modeling of \jpsi Yield Modifications in Proton
  (Deuteron) - Nucleus Collisions at High Energies}

\author{J. L. Nagle}
\affiliation{University of Colorado at Boulder}
\email{jamie.nagle@colorado.edu}

\author{A. D. Frawley}
\affiliation{Florida State University}
\email{afrawley@fsu.edu}

\author{L. A. Linden Levy}
\affiliation{Lawrence Livermore National Laboratory}
\email{lindenle@llnl.gov}

\author{M. G. Wysocki}
\affiliation{University of Colorado at Boulder}
\email{matthew.wysocki@colorado.edu}

\date{\today}

\begin{abstract}
Understanding the detailed production and hadronization mechanisms for
heavy quarkonia and their modification in a nuclear environment
presents one of the major challenges in QCD. Calculations including
nuclear-modified parton distribution functions (nPDFs) and fitting
of break-up cross sections (\sbr) as parameters have been successful at
describing many features of \jpsi modifications in
proton(deuteron)-nucleus collisions. In this paper, we extend these
calculations to explore different geometric dependencies of the
modifications and confront them with new experimental results from the
PHENIX experiment. We find that no combination of nPDFs and \sbr,
regardless of the nPDF parameter set and the assumed geometric dependence, can
simultaneously describe the entire rapidity and centrality dependence
of \jpsi modifications in \dau collisions at $\sqsn=200$ GeV. We also
compare the data with coherence calculations
and find them unable to describe the full rapidity and centrality
dependence as well. We discuss how these calculations might be extended and
further tested, in addition to discussing other physics mechanisms including
initial-state parton energy loss.
\end{abstract}


\maketitle

\section{Introduction}

In proton(deuteron)-nucleus collisions at the Relativistic Heavy Ion
Collider (RHIC), the nucleus is
extremely Lorentz-contracted and thus the
entire interaction and traversal of the nuclear target takes
place on a time scale of order 0.1 fm/$c$. Thus, one expects coherence
effects to play a significant role in the physics of particle
production and hadronization.   By studying heavy
quarkonia states, one can postulate that the initial hard production of a
\ccbar pair can be factorized from the later traversal of
that pair through the remainder of the nucleus. Many calculations
have utilized this factorized approach in trying to understand the
nuclear modification of \jpsi yields in proton(deuteron)-nucleus
reactions (for
example~\cite{Lourenco:2008sk,Adare:2007gn,Vogt:2004dh,Kharzeev:1996yx}). In
this factorized framework, modification of initial \ccbar pair production is
accounted for via nuclear-modified parton distribution functions (nPDFs).  
After being produced, the disassociation of charm pairs and
thus the additional reduction in the number of final-state \jpsi mesons is
accounted for with a simple breakup cross section ($\sigma_{br}$).  It is
interesting to test this picture to see if the experimental data requires
additional physics, including coherence effects and initial-state parton
energy loss.
This work presents an extension of this simple calculational framework
and investigation of some of the key underlying assumptions. 

\section{Calculation Details}

\begin{figure}[htb]
  \includegraphics[width=0.9\linewidth]{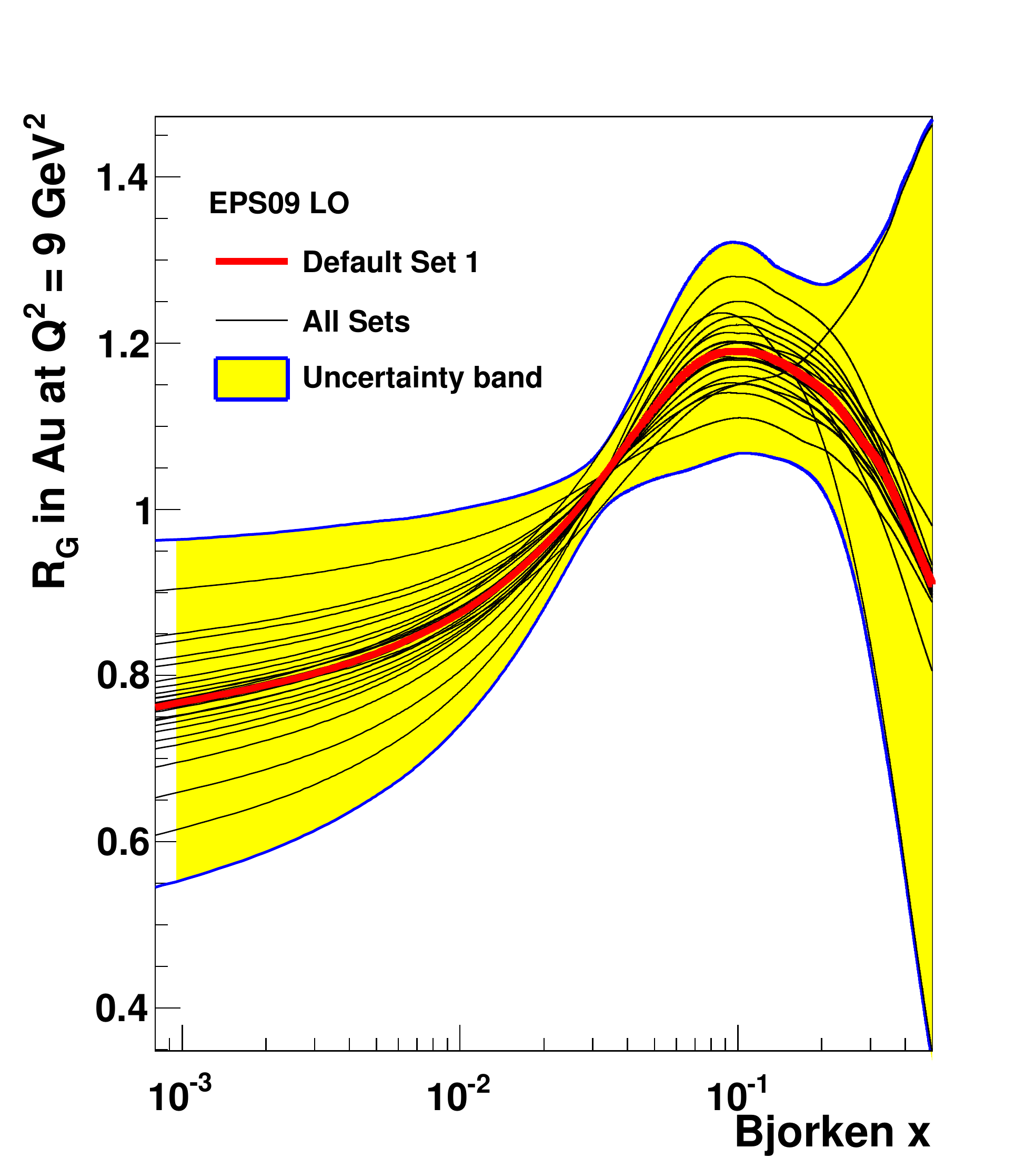} 
  \caption{
  	The gluon nuclear modification $R_{G}$ for the Au nucleus at the scale 
	$Q^{2}$ = 9 GeV$^2$ is shown for the EPS09 central value (labeled set 1) and for all 30
	error sets. The yellow shaded area is the overall uncertainty band calculated 
	from the error sets, representing a 90\% CL uncertainty. The parameter sets from 
	top to bottom at the lowest $x$ values are 16, 10, 7, 8, 13, 31, 26, 18, 15, 4, 3, 21, 23, 25, 28, 1, 29, 
	2, 24, 20, 22, 14, 5, 19, 27, 30, 6, 12, 9, 11, 17.  }
  \label{fig_eps09_sets}
\end{figure}

In this section we describe the inputs required for the calculation of
the nuclear modification factors (\rpa or \rda) for various nuclear
targets and centrality selections. First, the density of partons in
the nucleus is modified relative to the parton distribution function
(PDF) for nucleons.  This nuclear-modified PDF reflects the modified
parton density encountered by the projectile and results in a modified
number of hard scatterings that create \ccbar pairs from $g + g$, $q +
g$, and $q + \overline{q}$ interactions.  The state of the art
calculation is the EPS09 nPDF parameter set with uncertainties
represented by 31 different Hessian basis parameterizations as
detailed in~\cite{Eskola:2009uj}.  Because \jpsi production at high energies
is dominated by interactions between gluons, we will consider $g+g$ interactions 
only in the calculations in this paper. Figure~\ref{fig_eps09_sets} shows
the EPS09 gluon modification $R_{G}$ at a $Q^{2}$ = 9 GeV$^2$, the
appropriate scale for production of the \jpsi. It can be seen that the
nPDFs are not well constrained by experimental data, particularly the
low-$x$ gluon distributions which dominate the \jpsi production
probability at forward rapidity at RHIC energies.

The second main effect is that after creating the \ccbar pair in the initial
state (often referred to as the \jpsi precursor, since the hadronic
state is expected to take of order 0.3 fm/$c$ to form), the pair may
break up or be de-correlated while traversing the remaining portion
of the nucleus. This second effect is often included by assuming
a fixed cross section \sbr for the breakup of the pair. We note that
this effect is also often termed absorption, though this nomenclature
can be misleading since the charm pair still exists, but is no
longer able to form a final-state \jpsi meson. Currently there is no
fundamental description of the hadronization process for the \jpsi
meson that agrees with all the available experimental
data~\cite{Brodsky:2009cf,Lansberg:2008gk}. The lack of such a
theory for the dynamics of hadronization means one has no {\it ab
  initio} calculations of this precursor-nucleon cross section and its
dependence on the relative velocity between the pair and the target
nucleons. In most works, the value of \sbr is assumed to be 
independent of the \jpsi rapidity for a given \sqsn, and is determined from fits to the
experimental data~\cite{Lourenco:2010zza}.

\begin{figure}[htb]
  \includegraphics[width=0.9\linewidth]{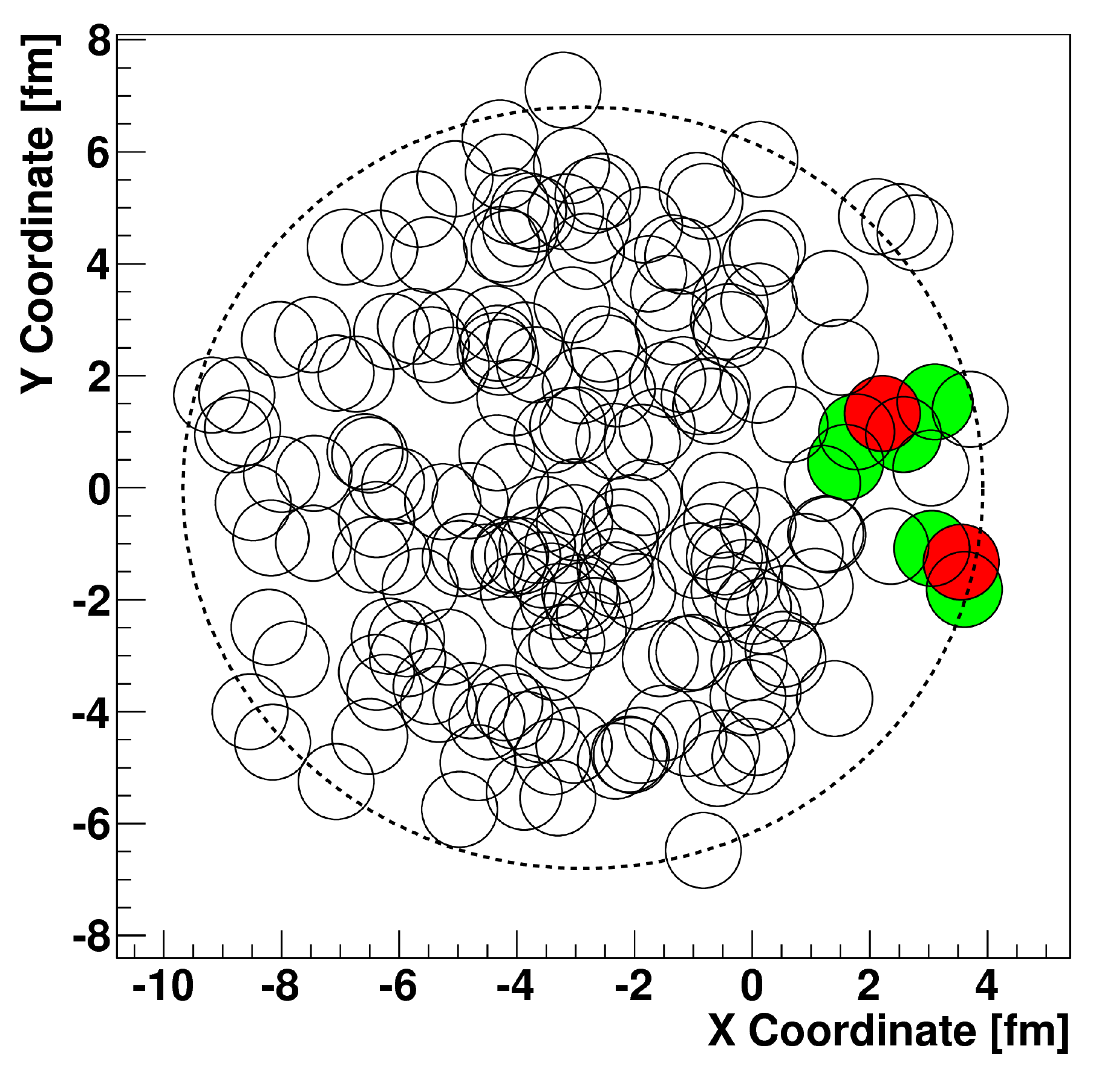} 
  \caption{Monte Carlo Glauber event display in the transverse (x-y) plane.
    Open black circles are the positions of the gold nucleus nucleons.
    Red filled circles are the positions of the two nucleons from the
    deuteron. Green filled circles are the positions of the gold
    nucleus nucleons which suffer at least one binary collision. 
    The dashed black circle represents gold nucleus extent from the
    Woods-Saxon parameterization.}
  \label{fig_dAu_display}
\end{figure}

 We employ a Monte Carlo Glauber model~\cite{Miller:2007ri} where the
nucleons are randomly given spatial distributions within the deuteron
based on the Hulthen wave function and for the gold nucleus based on a
Woods-Saxon distribution with parameters R=6.38 fm and a=0.54
fm~\cite{Adare:2007gn}. Individual \dau collisions at $\sqsn=200$ GeV
are simulated by randomly selecting an event impact parameter ($b$) and determining if
any pair of nucleons collide using an inelastic cross section of $\sigma = 42$ mb. 
One example event is shown in
Figure~\ref{fig_dAu_display}, where the open circles are the positions
of the gold nucleus nucleons in the transverse plane, the red filled
circles are the positions of the two nucleons from the deuteron, and
the green filled circles are the gold nucleus nucleons which suffered
a binary collision. 
Each binary collision between a red circle (deuteron nucleon)
and a green circle (gold nucleon) has a probability to produce a
\ccbar pair. This probability is modified from proton-proton
collisions according to the afore-mentioned nPDFs.

The EPS09 nPDF parameterization, as well as other nPDF
parameterizations, are predominantly determined from deep inelastic
scattering experiments and minimum bias $p+A$ reactions producing
Drell-Yan pairs~\cite{Eskola:2009uj}. In such experiments there is no measure of the impact
parameter or transverse distance within the nucleus for the interaction and 
therefore the geometric dependence of the nPDF modification is not constrained.
One expects a significant geometric dependence in the nPDF modification, with a 
stronger modification near the center of the nucleus where the density of nucleons is the
largest. In a simple picture, the partons inside the nucleons at low-$x$
have wave functions that are longer in the longitudinal direction than
the Lorentz contracted nucleus.  Thus the nPDF modification depends
on the density of overlapping nucleons as shown in the transverse
plane in Figure~\ref{fig_dAu_display}. However, there is no such
Lorentz contraction in the transverse direction, and the parton wavefunction
extent in this plane is of order 1 fm. Therefore, the largest
nuclear effect would be expected where the density, and thus the
longitudinal overlap is largest, near the center of the nucleus and
should decrease as one moves out toward the periphery.

In~\cite{Klein:2003dj}, the nPDF modification is postulated to be
linearly proportional to the density-weighted longitudinal thickness
of the nucleus at the transverse position of the binary collision, as written
below:
\begin{equation}
\mrt = 1.0 - a \lrt
\label{eq:mrt}
\end{equation}
where $\lrt = {{1}\over{\rho_{0}}} \int dz \rho(z,\rt)$ is the
density-weighted longitudinal thickness and $\rho_{0}$ is the density
at the center of the nucleus. In Figure~\ref{fig_dAu_display}, each
green circle is a transverse distance \rt from the center of the gold
nucleus, and one can then determine the average local thickness \lrt based on
the Woods-Saxon parameterization.

As shown in Figure~\ref{fig_dAu_display}, there are significant
fluctuations in the thickness \lrt due to the randomly selected
spatial locations of the nucleons in the gold nucleus at the time of
the collision. In fact, the inclusion of such fluctuations has proven
to be crucial in modeling the initial conditions in heavy ion
reactions (see for
example~\cite{Alver:2010gr,Sorensen:2010zq,Nagle:2009ip,Drescher:2007cd,Alver:2006wh}).
In order to incorporate these fluctuations, we calculate the number of
target nucleons around the struck nucleon (green circle) within a
transverse radius $\rtub = 2 \times R_{nucleon}$, where $R_{nucleon}$
is taken to be the charge radius of the proton, 0.87
fm~\cite{Nakamura:2010zzi}.  The average number of nucleons within
this cylinder \ntub is $\pi \rtub^{2} \rho_{0}$ times \lrt, and this
proportionality constant can be absorbed into the parameter $a$ in
Eqn.~\ref{eq:mrt}.  The exact choice of \rtub is somewhat arbitrary,
but reasonable changes in the value do not significantly change the
results shown in this paper.

\begin{figure}[htb]
  \includegraphics[width=0.9\linewidth]{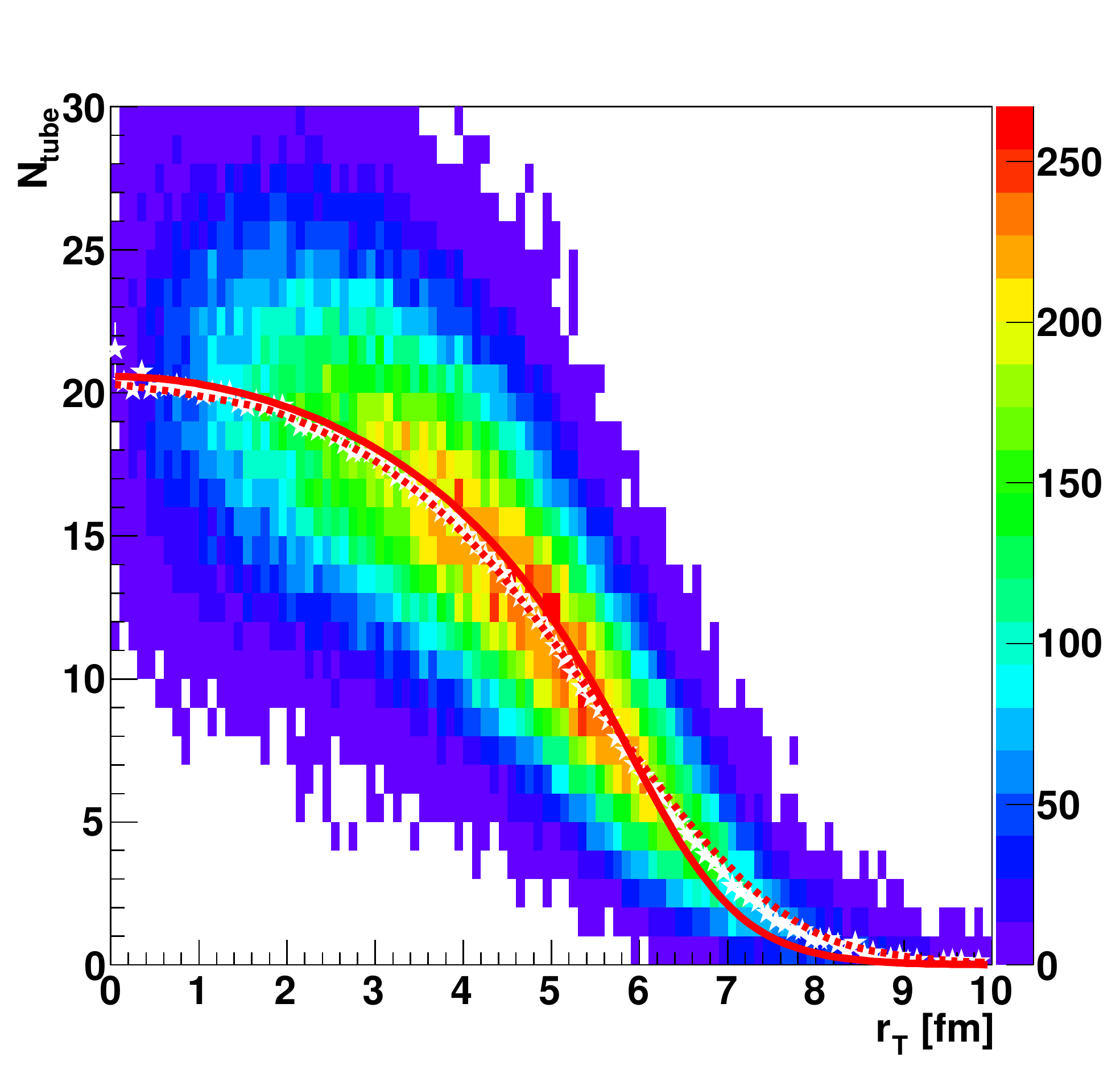} 
  \caption{
    Shown is the Monte Carlo Glauber result for the \ntub
    distribution as a function of \rt. The mean values of \ntub as a
    function of \rt are shown as white points. The analytic
    calculation of the average value of \lrt from the Woods-Saxon
    parameterization, rescaled to the \ntub value, is shown as the solid red curve. 
    Smearing the analytic calculation around \rt by the tube radius \rtub 
    is shown as the red dashed curve.
}
  \label{fig_lambda_fluc}
\end{figure}

In Figure~\ref{fig_lambda_fluc} the resulting distribution of \ntub
values as a function of \rt is shown. For collisions in the middle of
the nucleus ($\rt \approx 0$)  $\left< \ntub \right> \approx 20$ and the RMS
$\approx 5$. 
Using the analytic calculation of the average \lrt from the
Woods-Saxon parameterization times $\pi \rtub^{2} \rho_{0}$ yields
the solid red curve. The difference between the two is because the
$\ntub(\rt)$ calculation includes the density averaged over the tube
radius \rtub. If we smear the analytic calculation around \rt by the
tube radius \rtub, we obtain the red dashed curve, which now shows
much better agreement.

\begin{figure}[htb]
  \includegraphics[width=0.9\linewidth]{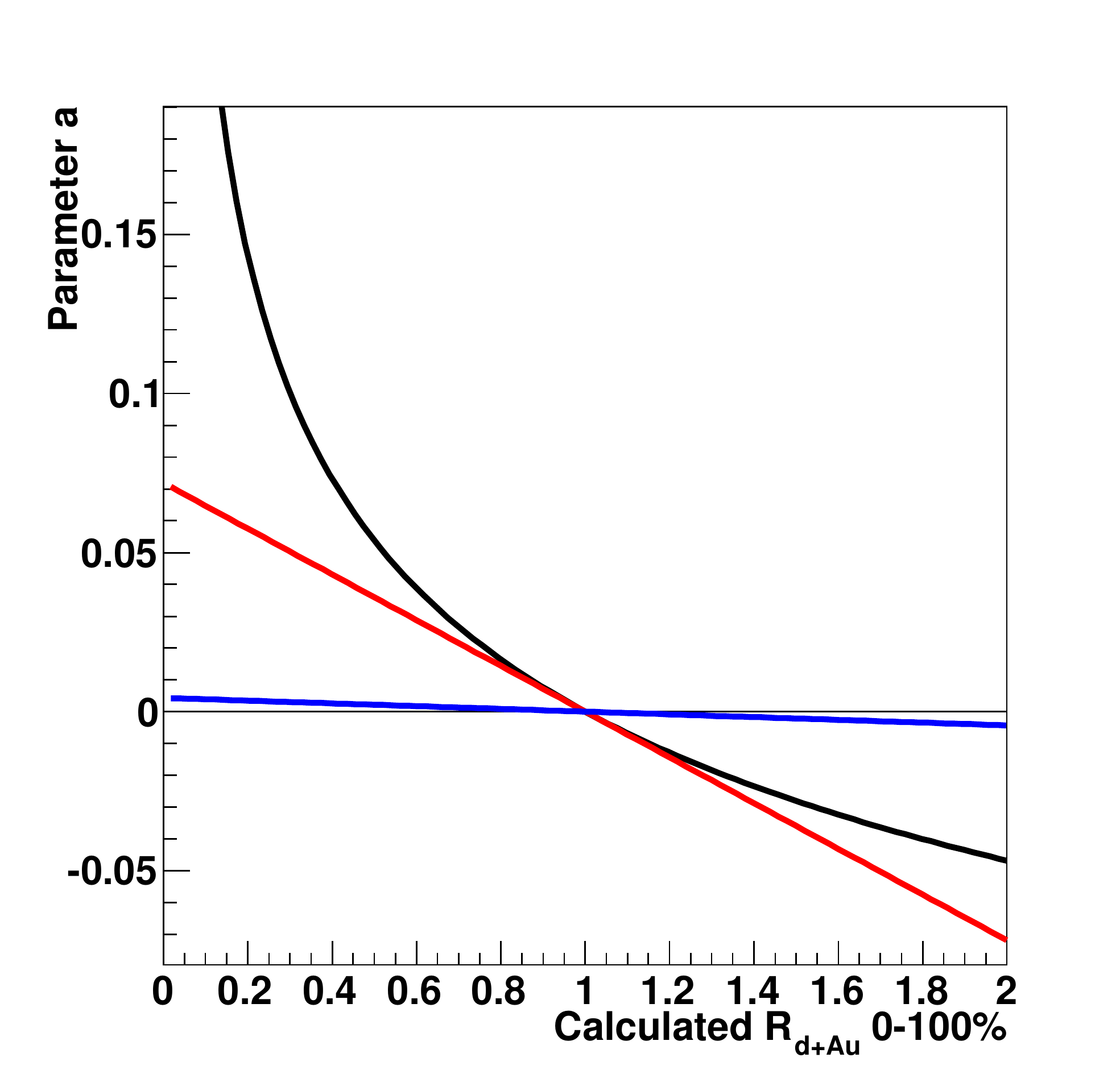} 
  \caption{Shown is the parameter
    $a$ as a function of the \rdauave modification that is obtained.
    The red, black, and blue curves correspond to the linear, exponential,
    and quadratic geometric dependence cases respectively.
  }
  \label{fig_parametera}
\end{figure}

In our calculations, we take these fluctuations into account by
utilizing \ntub instead of the average \lrt, as shown in the equation below:
\begin{equation}
\mrt = 1.0 - a \ntub(\rt)
\end{equation}
There is a direct relationship between the parameter $a$ and the
$\langle M\rangle$, the modification averaged over all nuclear
geometries, which is equivalent to \rdauave.  This relationship is
determined by averaging $M$ over the \rt distribution for unbiased
collisions as determined from the Monte Carlo Glauber (shown in Figure
3 of the PHENIX publication~\cite{Adare:2010fn}).
The results are shown as the red curve in Figure~\ref{fig_parametera}. 
We also consider two other geometric dependencies for the
nPDF, referred to as exponential and quadratic.
\begin{eqnarray}
\mrt = Exp\left( - a \ntub(\rt)\right)&\\
\mrt = 1.0 - a \left[\ntub(\rt)\right]^{2} &
\end{eqnarray}
The same procedure is applied, and the dependence of the parameter $a$
is also shown in Figure~\ref{fig_parametera} as the black (blue) curve for
the exponential (quadratic) case.

\begin{figure*}[!tb]
  \centering
  \includegraphics[width=0.9\linewidth]{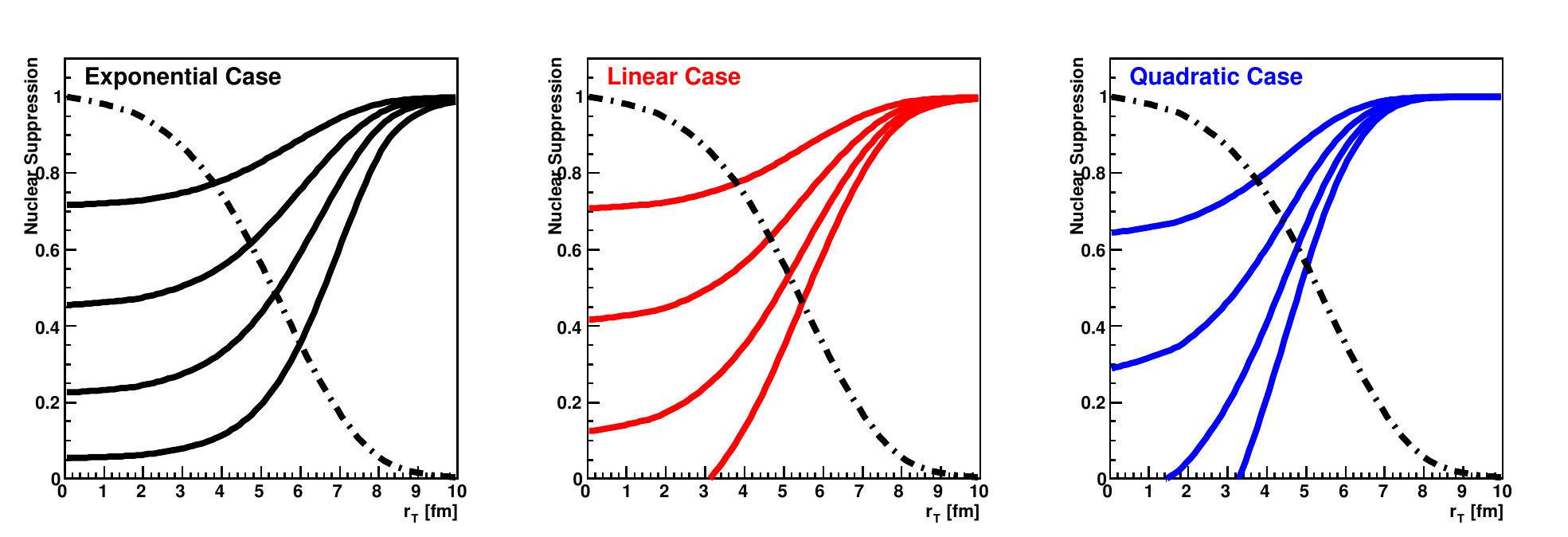} 
  \caption{Shown are the modification dependencies \mrt for the three geometry cases
    (exponential, linear, and quadratic) assuming four different
    average \rdauave values -- 0.8, 0.6, 0.4, 0.2 (corresponding to the top
    to bottom curves in each case). The dot-dashed
    black line is the shape of the \ntub distribution versus \rt,
    normalized to one at $\rt=0$ for reference.
  }
  \label{fig_rt_dep}
\end{figure*}

We examine the modification \mrt for each of these cases.  Shown in
Figure~\ref{fig_rt_dep} (left panel) is \mrt for the exponential
case. The four solid black curves correspond to geometry averaged
modifications of $\left< M \right>$ = 0.8, 0.6, 0.4, 0.2 for the top
to bottom curves. The dot-dashed black line is the shape of the mean
of the \ntub distribution versus \rt, normalized to one at $\rt=0$ for
reference. The middle (right) panel shows the same results for the
linear (quadratic) case. As a consequence of the linear and quadratic
functional forms, \mrt has negative values for small \rt when the
geometry-averaged modification is less than 0.4 (0.6) for the linear
(quadratic) cases. This unphysical result can be removed by forcing
the modification to be positive-definite, but then one has to
recalculate the corresponding $a$ parameters.
In the analysis presented in this paper, the modification values from the 
nPDF do not typically reach these low
values and thus we have not had to recalculate the results. 
However careful attention to this problem
will be crucial for cases with larger modifications (for example more
forward rapidity \jpsi measurements or very low-$x$ measurements at the
Large Hadron Collider in $p+A$ and $A+A$).

There are two final ingredients needed to map the nPDF modifications
onto the final state \jpsi suppression. These are the distribution of
Bjorken \x{2} and \qsq for the parton-parton processes that contribute
to the \jpsi production and the mixture of $g+g$, $g+q$, and $q + \overline{q}$ 
processes. The simplest relationship between the \jpsi
and partonic kinematics arises under the assumption that the
production is a $2 \rightarrow 1$ process, for example $g + g
\rightarrow \jpsi$. Invoking conservation of energy and momentum the
production of a \jpsi with $\pt=0$ GeV/$c$ results in the following
relationship between \jpsi rapidity ($y$) and parton momentum fraction
(\x{2}).
\begin{equation}
\x{2} = \frac{M_{\jpsi}}{\sqsn} e^{-y}
\end{equation}
This $2 \rightarrow 1$ process is actually forbidden by angular momentum conservation, but may approximate the
correct kinematics at low \pt, or in a color evaporation picture where
soft gluon emission does not significantly modify the exact
correlation of \x{2} and $y$. It has been pointed out that with a
more detailed understanding of the subprocesses that contribute to
\jpsi production, one can utilize a more exact map of \x{2} and \qsq
to the final \jpsi as a function of rapidity and
\pt~\cite{Rakotozafindrabe:2010su,Ferreiro:2009ur,Ferreiro:2008wc}. 
The authors utilize the following relation between $x_{1}$ and $x_{2}$ that requires
a full modeling of the cross section dependencies:
\begin{equation}
\x{2} = \frac{ \x{1} \sqrt{ \pt^{2} + M^{2} } \sqsn e^{-y} - M^{2} }{\sqsn \left( \sqsn x_{1} - \sqrt{\pt^{2} + M^{2} } e^{y} \right) }
\label{eqn22}
\end{equation}
This relation is exact for a $2 \rightarrow 2$ process where one
outgoing particle is an on-shell \jpsi and the other particle is massless.

\begin{figure}[htb]
  \includegraphics[width=0.9\linewidth]{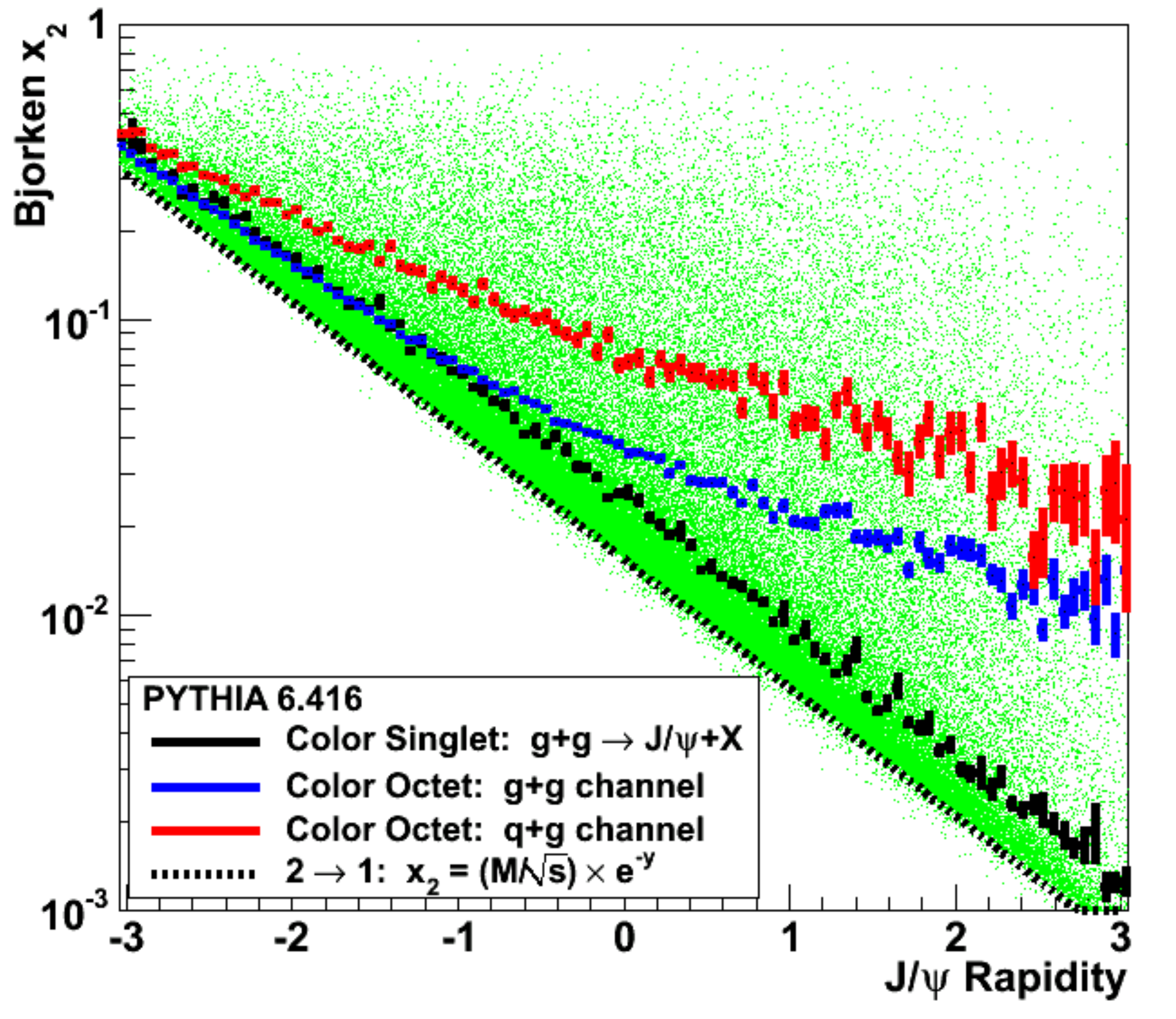} 
  \includegraphics[width=0.9\linewidth]{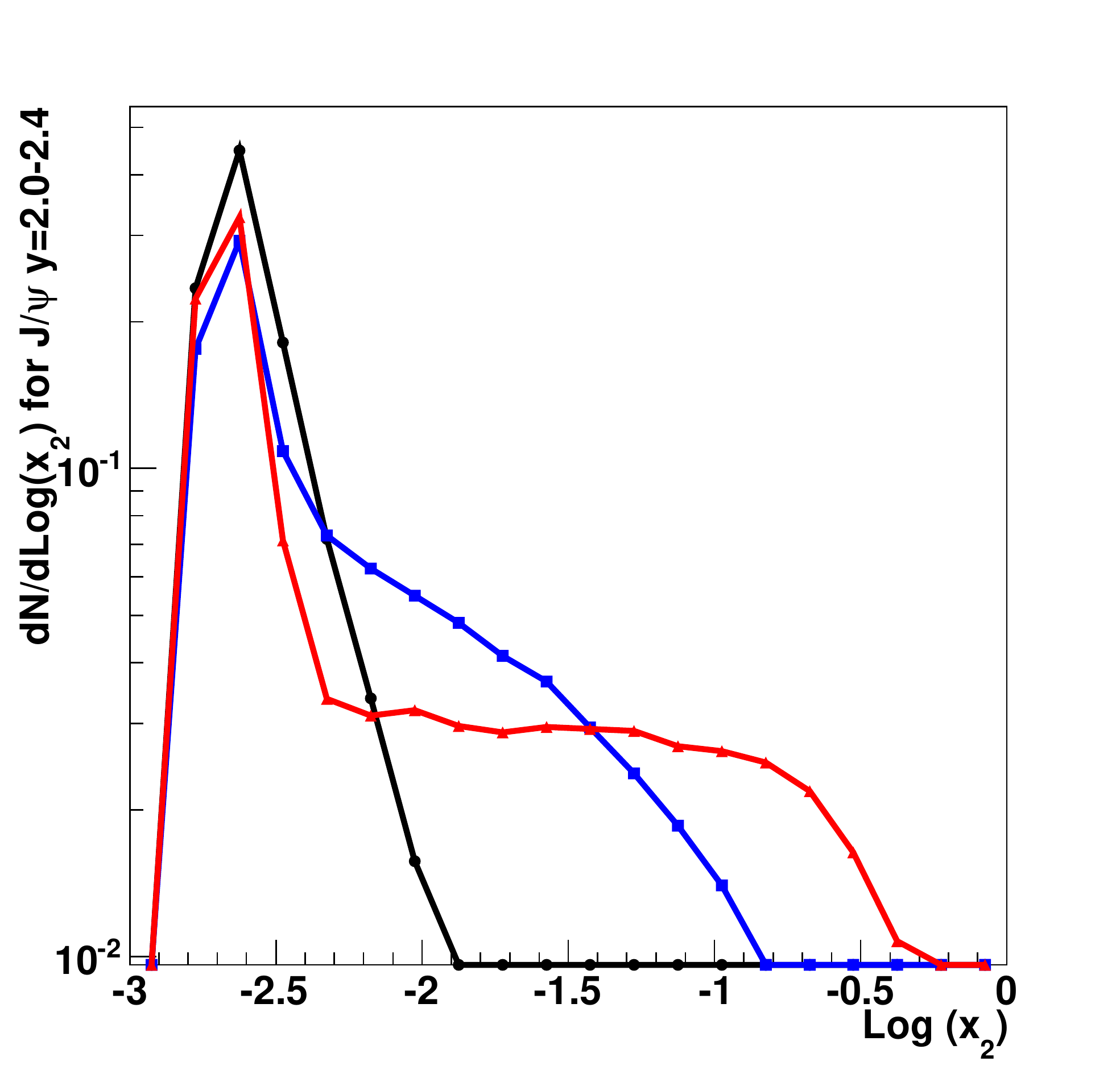} 
  \caption{(Upper panel)  The kinematic correlations between
    the \jpsi rapidity and Bjorken $x_{2}$ for the $2 \rightarrow 1$
    process compared with a scatter-plot showing calculations from PYTHIA 6.416 
    with the NRQCD setting for \jpsi production. The $\left< x_{2} \right>$ values are shown
    for three different production mechanisms within PYTHIA 6.416.
    (Lower panel)  Shown are the $x_2$ distributions for \jpsi produced
    with rapidity $2.0 < y < 2.2$ for the three different PYTHIA production
    mechanisms (black for $g+g$ color singlet, blue for $g+g$ color octet,
    and red for $q+g$ color octet).
    }
\label{fig_x_mapping}
\end{figure}

Shown in Figure~\ref{fig_x_mapping} (upper panel) is the correlation of
Bjorken \x{2} with the \jpsi rapidity in the $2 \rightarrow 1$ case.
This is compared with a scatter-plot showing calculation results from PYTHIA 6.416~\cite{Sjostrand:2006za} with the
NRQCD setting for \jpsi production.  As expected, the \x{2} values
for a given \jpsi rapidity are shifted to larger values. Since the
 \jpsi $\langle\pt\rangle\approx$ 2.2 GeV/$c$, there must be a balancing
particle(s), which requires larger available energy. Also, the
emission of a balancing gluon, for example, will smear the rapidity of
the \jpsi relative to the $2 \rightarrow 1$ calculation. 
Also shown are the $\left< x_{2} \right>$ values as a function of rapidity
for the PYTHIA $g+g$ color singlet channel (black), $g+g$ color octet channel (blue),
and the $q+g$ color octet channel (red).  The mean $x_{2}$ value can be misleading since
it may have a large influence from a small fraction of high-$x$ events.  Thus,
in the lower panel we show the log($x_{2}$) distribution for the \jpsi rapidity $2.0 < y < 2.4$
for the three different contributions.  The majority of processes for this rapidity
involve $x_{2} \approx 0.002$, but with a more significant high-$x$ tail in the octet
cases.  Note that the underlying PYTHIA production does not obey the $2 \rightarrow 2$ kinematics
of Eqn.~\ref{eqn22}, since there is initial-state $k_{T}$ and many of the
octet production channels involve more than two final-state particles.




We now incorporate all of the following items: (1) Monte Carlo
Glauber, (2) deuteron and gold nuclear geometries, (3) EPS09 nPDF parameter set,
(4) geometric dependence assumption for nPDF (i.e. linear, quadratic,
exponential), (5) kinematic mapping ($g$, $q$, $\overline{q}$,
\x{2}, \qsq $\rightarrow$ \jpsi $y$, \pt). We then add the second
factorized part of the calculation for the \sbr by checking all
nucleons on the back side of the nucleus (i.e. $z_{nucleon} >
z_{binary}$) for whether the \ccbar pair breaks up. Note
that although the calculations are factorized, the results are
auto-correlated by the geometry. For example, a binary collision
occurring near $\rt = 0$ has a larger nPDF modification and also a
larger probability for breakup. These auto-correlations are
important to account for, and have been previously explored in terms of $k_T$ kicks
broadening the \pt distribution~\cite{Nagle:1999ms,Kharzeev:1997ry}.

Two additional benefits of this Monte Carlo Glauber approach with full
fluctuations are that we can model the exact PHENIX experimental \dau
centrality selection event-by-event and that we never project
onto an averaged quantity ($eg.$ the average impact parameter for each
centrality class) and then calculate the modification for
that average quantity.

\begin{figure}[htb]
  \includegraphics[width=0.95\linewidth]{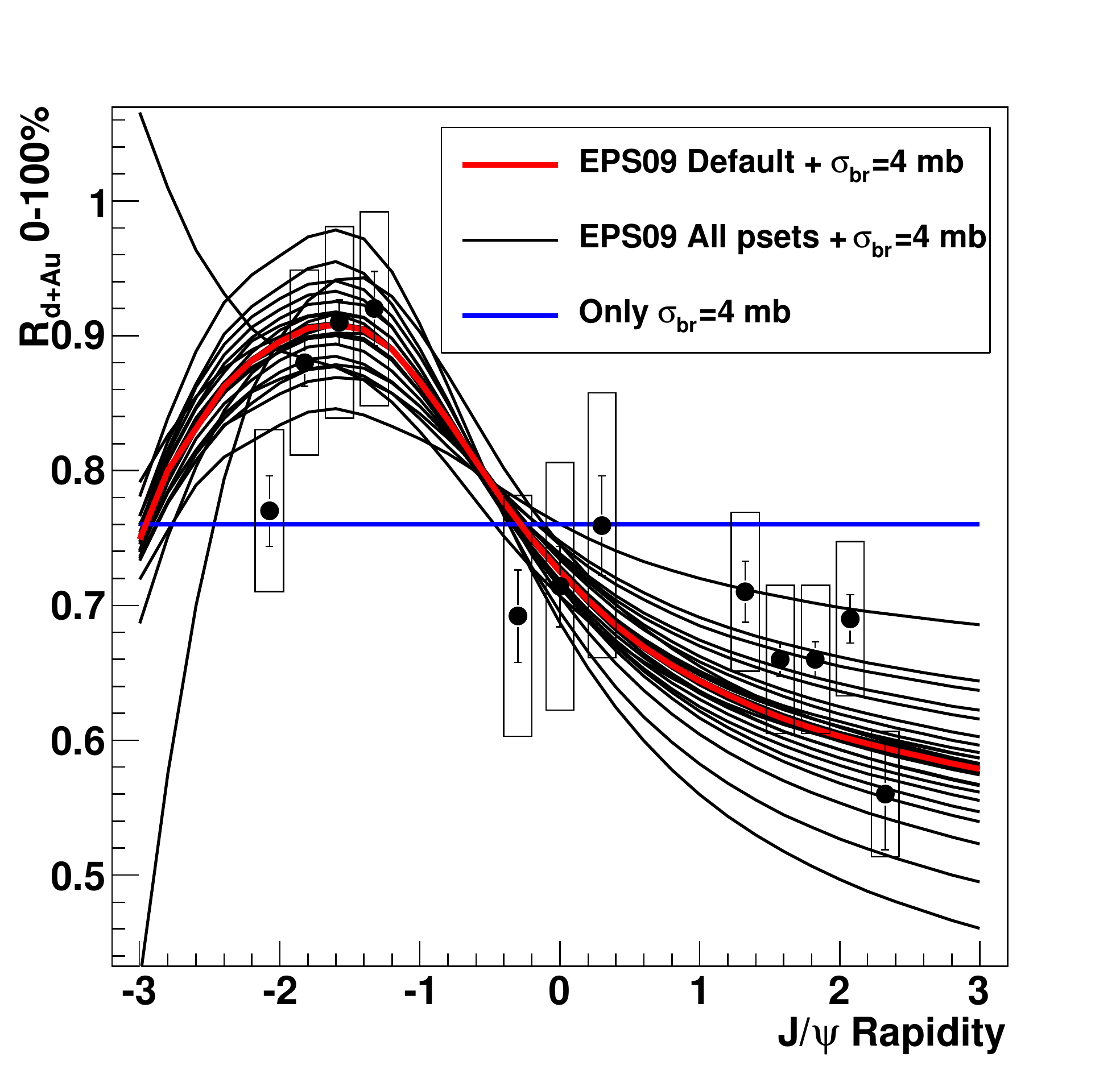} 
  \caption{
    Shown is the \jpsi nuclear modification factor \rdau for 0-100\% interaction
    centrality as a function of rapidity.  The calculations include the
    EPS09 nPDFs with the linear geometric dependence and $2 \rightarrow 1$ kinematics.
    The yellow band shows the limits from all 31 EPS09 nPDF variations, but should not
    be interpreted as a one-standard deviation uncertainty band.
    The PHENIX experimental data are shown as black points.  The lines are the
    point-to-point uncorrelated uncertainties and the boxes are the point-to-point
    correlated systematic uncertainties.  Not shown is the additional $\pm 7.8$\% 
    global scale uncertainty.  
}
  \label{fig_example}
\end{figure}

\section{Calculation Results}

Putting all these pieces together, we show an example
calculation in Figure~\ref{fig_example} of the \jpsi nuclear
modification \rdau as a function of rapidity for 0-100\% of the \dau
inelastic cross section. In this example, we utilize the $2
\rightarrow 1$ exact process mapping and the linear
geometric dependence of the nPDFs. We show the default EPS09 result with a $\sbr = 4$ mb
(red curve), all 30 other variations for EPS09 with a $\sbr = 4$ mb
(black curves), and a calculation assuming no nPDF modification with $\sbr = 4$ mb
(blue curve).  
The PHENIX experiment has recently reported high statistics \jpsi \dau
at $\sqsn= 200$ GeV nuclear modification factors as a function of
rapidity~\cite{Adare:2010fn} which are also shown in Figure~\ref{fig_example}.
Within systematic uncertainties from the experimental data and theoretical
calculation, reasonable agreement is obtained.

We emphasize that this calculation utilized the $2 \rightarrow 1$ kinematics.  
We have performed the same calculation using the various PYTHIA
kinematics and find only a very modest decrease (smoothing out) of the 
rapidity dependence.  The rapidity dependence in the calculation comes entirely from
the nPDF dependence on $x_{2}$ and $Q^{2}$.  Thus, utilizing the PYTHIA
kinematics leads to a slight blurring of this relation and a general
shift to larger $x_{2}$ values, as expected from Figure~\ref{fig_x_mapping}.  This flattening of the \rdau versus rapidity
goes in the opposite direction of the experimental data; however, the
uncertainties in the nPDFs and other physics do not allow for any conclusion
about the underlying production process.


\begin{figure}[t]
  \includegraphics[width=0.9\linewidth]{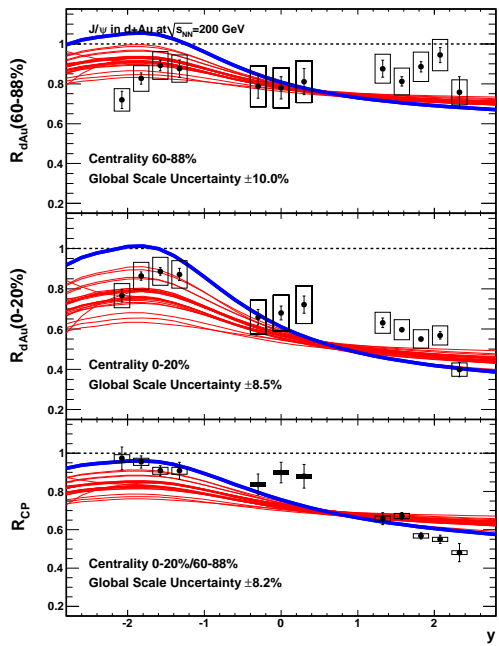} 
  \caption{Comparison of PHENIX data with calculations (red curves) using the 31 EPS09 parameter 
  sets with linear nuclear thickness dependence. Each EPS09 parameter set is shown for its own best 
  fit $\sbr$ value. The blue curve shows the best fit for the best set, obtained with EPS09 set 17 and $\sbr = 3.2$ mb.
}
  \label{fig_rcp_data_line}
\end{figure}

\begin{figure}[t]
  \includegraphics[width=0.9\linewidth]{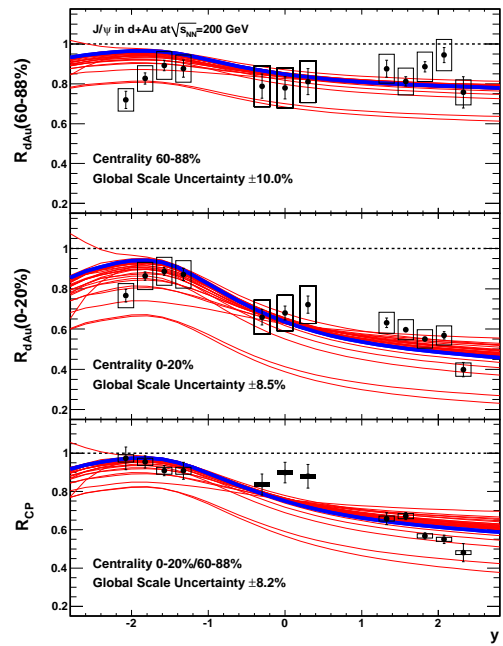} 
  \caption{Comparison of PHENIX data with calculations (red curves) using the 31 EPS09 parameter 
  sets with quadratic nuclear thickness dependence. Each EPS09 parameter set is shown for its own best 
  fit $\sbr$ value. The blue curve shows the best fit for the best set, obtained with EPS09 set 30 and $\sbr = 3.4$ mb.
}
  \label{fig_rcp_data_quad}
\end{figure}

Shown in Figure~\ref{fig_rcp_data_line}
are the PHENIX experimental results for \rdau for peripheral events
60-88\% (top panel), \rdau for central events 0-20\% (middle panel),
and the ratio between them \rcp 0-20\% / 60-88\% (lower panel). Note
that the significant systematic uncertainties that can modify the
rapidity dependence of the modification, referred to by PHENIX as
type-B systematics, largely cancel in the \rcp ratio. 

We utilize the \rt distributions for each centrality class shown in Figure 3 of the PHENIX
publication~\cite{Adare:2010fn} to compute the expected modification
in each centrality.
There are many different statistical fits one can perform between the experimental
data and our theoretical calculations.  In this
case, we perform a modified-$\chi^{2}$ ($\tilde{\chi}^{2}$) fit to just the \rcp data (which
provides by far the best constraint on the rapidity dependence). The
$\tilde{\chi}^{2}$ fit method that accounts for both statistical and systematic
uncertainties is detailed in~\cite{Adare:2008cg}. 

We have chosen again to utilize the $2 \rightarrow 1$ kinematics and
consider the linear, quadratic, and exponential geometric dependencies
for the nPDFs.  Shown in Figure~\ref{fig_rcp_data_line} are the results 
for the linear geometric dependence.
Each red curve represents one of the 31 EPS09
nPDF parameter sets and the best fit \sbr value for that parameter set ($ie.$ minimum 
$\tilde{\chi}^{2}$) to the \rcp data. The calculation corresponding to that
red curve is then also shown for \rdau peripheral and central in the
upper and middle panels, respectively.  The blue curve represents the
best fit overall for all combinations for EPS09 parameter sets and
$\sbr$ values -- corresponding to EPS09 nPDF parameter set 17 and a $\sbr = 3.2$ mb. 
However, even this best fit has a $\tilde{\chi}^{2}$ = 41.5 
which corresponds to an extremely poor fit (i.e. probability less than $10^{-4}$).  
The result with the geometric nPDF exponential case are similar
with the best fit from EPS09 nPDF parameter set 17 and $\sbr = 4.2$ mb and a
poor $\tilde{\chi}^{2} =$ 50.6.

Shown in Figure~\ref{fig_rcp_data_quad} are the results for the quadratic
geometric dependence.
In this case the best fit corresponds to EPS09 parameter set 30 and $\sbr = 3.4$ mb. Again,
even this best fit has a $\tilde{\chi}^{2} =46.3$ which corresponds
to an extremely poor fit.
One notable feature that may appear counter-intuitive is
that for some EPS09 nPDF parameter sets, the best fit shown by the red curve
appears very far below the \rcp data points. Because the rapidity
shape is so poorly matched, it is possible that a better fit is
obtained with the $\tilde{\chi}^{2}$ under the assumption that the
global scale uncertainty of 8.2\% has a three standard deviation
fluctuation low.


\section{New Geometric Constraints}

\begin{figure}[t]
  \includegraphics[width=0.95\linewidth]{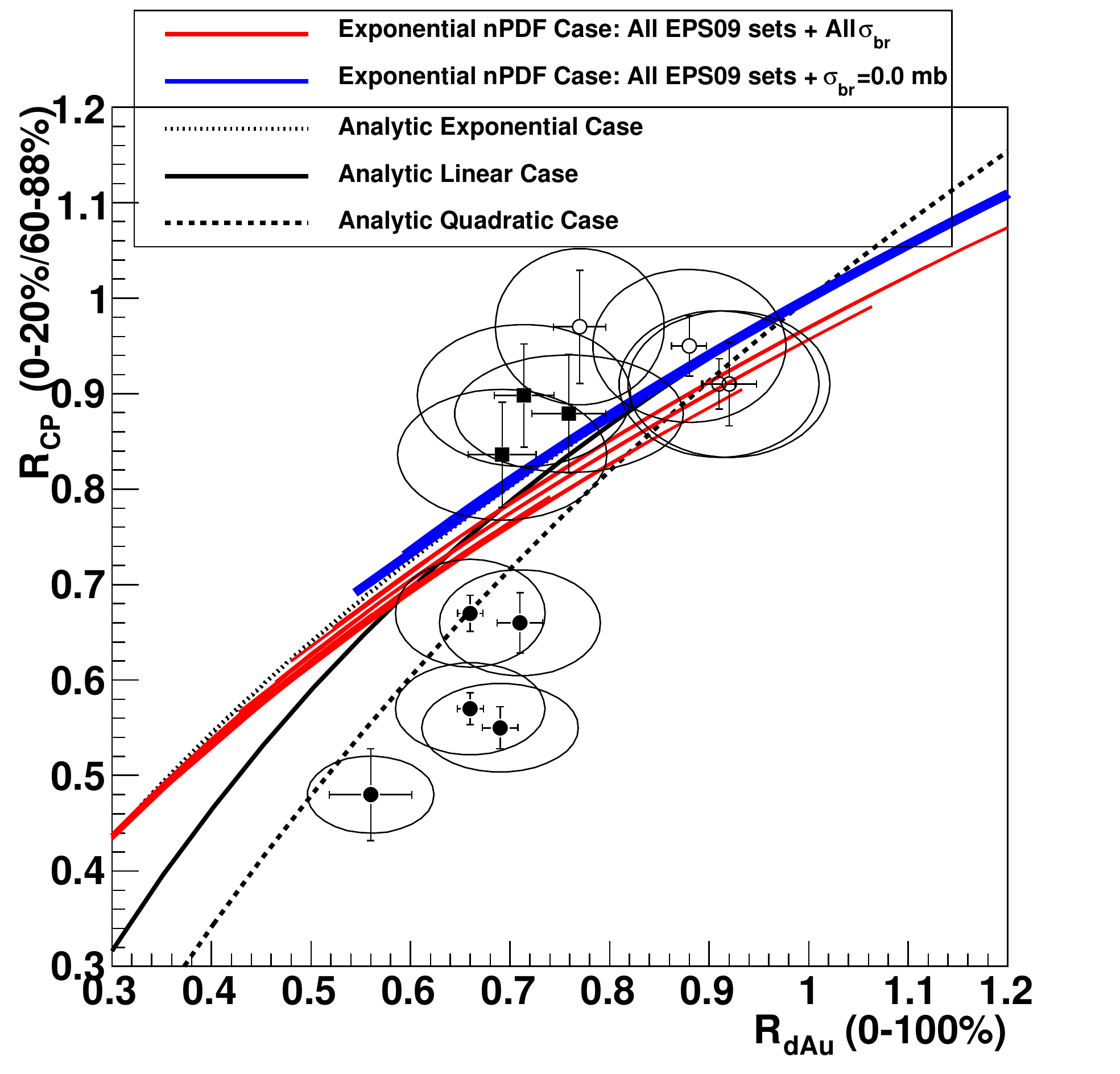} 
  \caption{The points are the PHENIX \jpsi \rcp versus \rdau.  The ellipses are the one-standard deviation contours for the systematic uncertainties.
    The open circles, closed squares, and closed circles  are from backward, mid, and forward rapidity respectively.
    The black curves are analytic calculations assuming a purely exponential, linear, and quadratic
  geometric dependence for the nuclear modification~\cite{Adare:2010fn}. The blue lines are the full calculation,
  using an exponential thickness dependence for the shadowing, for all EPS09 nPDF parameter sets
  with $\sbr = 0$. The red lines are the full calculation for all EPS09 nPDF parameter sets with all $\sbr$ values.
}
  \label{fig_expcase}
\end{figure}

\begin{figure}[t]
  \includegraphics[width=0.9\linewidth]{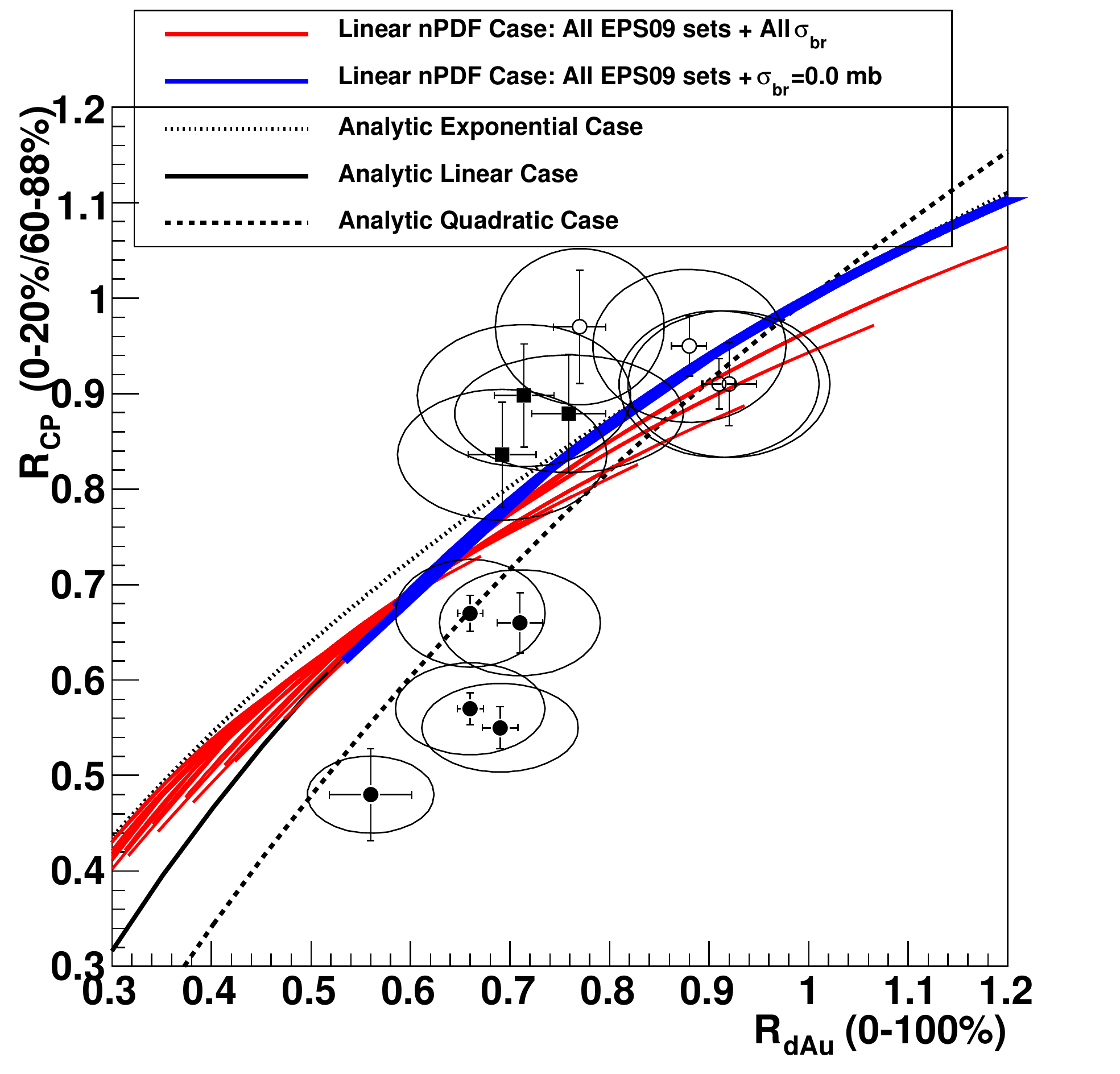} 
  \caption{The points are the PHENIX \jpsi \rcp versus \rdau.  The ellipses are the one-standard deviation contours for the systematic uncertainties.
    The open circles, closed squares, and closed circles  are from backward, mid, and forward rapidity respectively.
    The black curves are analytic calculations assuming a purely exponential, linear, and quadratic
  geometric dependence for the nuclear modification~\cite{Adare:2010fn}. The blue lines are the full calculation,
  using a linear thickness dependence for the shadowing, for all EPS09 nPDF parameter sets
  with $\sbr = 0$. The red lines are the full calculation for all EPS09 nPDF parameter sets with all $\sbr$ values.
}
  \label{fig_lincase}
\end{figure}

\begin{figure}[t]
  \includegraphics[width=0.9\linewidth]{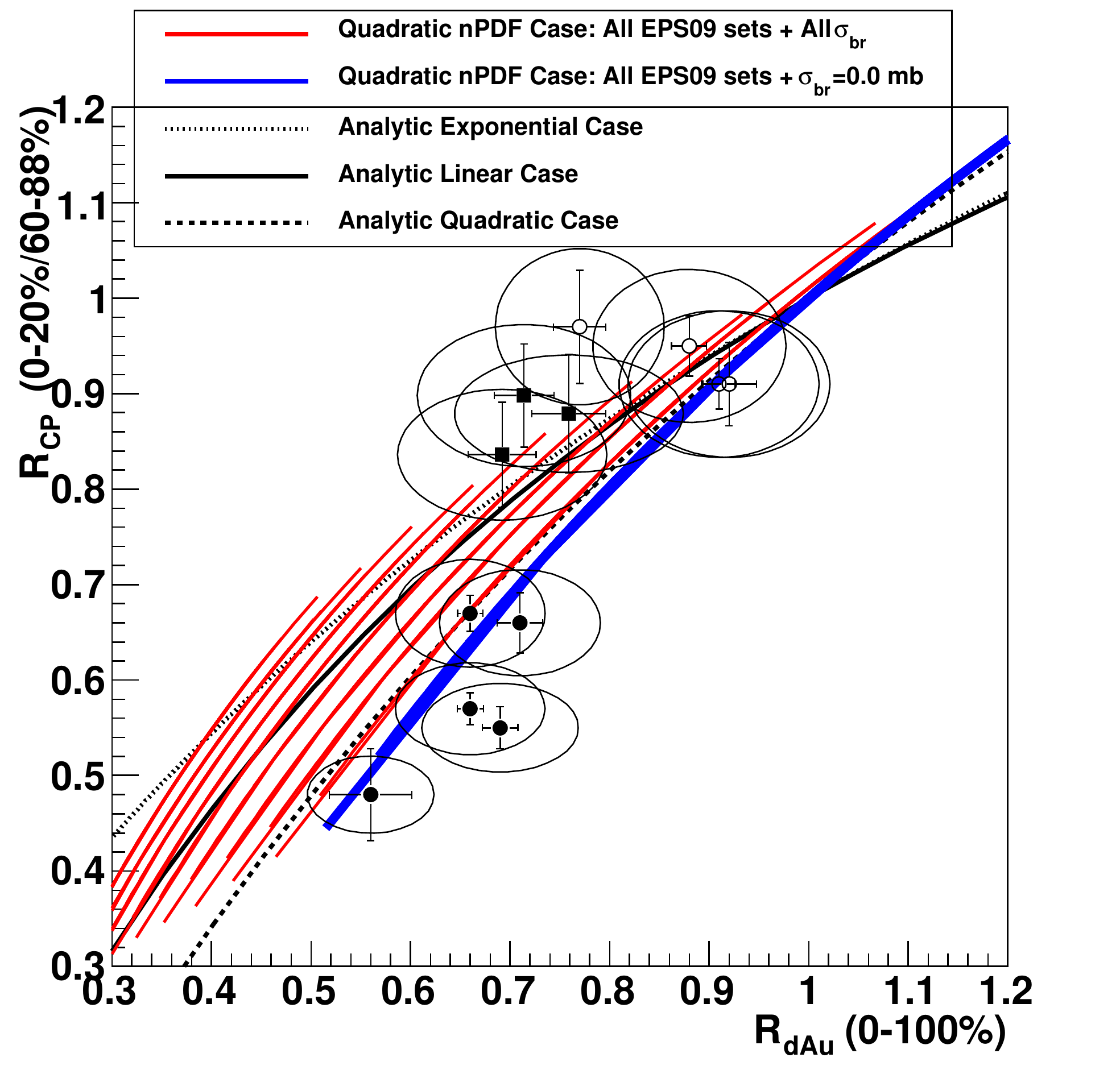} 
  \caption{The points are the PHENIX \jpsi \rcp versus \rdau.  The ellipses are the one-standard deviation contours for the systematic uncertainties.
    The open circles, closed squares, and closed circles  are from backward, mid, and forward rapidity respectively.
  The black curves are analytic calculations assuming a purely exponential, linear, and quadratic
  geometric dependence for the nuclear modification~\cite{Adare:2010fn}. The blue lines are the full calculation,
  using a quadratic thickness dependence for the shadowing, for all EPS09 nPDF parameter sets
  with $\sbr = 0$. The red lines are the full calculation for all EPS09 nPDF parameter sets with all $\sbr$ values.
}
  \label{fig_quadcase}
\end{figure}

In~\cite{Adare:2010fn}, the PHENIX collaboration presented a new way of using the
experimental data to test and constrain the geometric dependence of
the combined nuclear effects. By plotting the \rcp value (0-20\%/60-88\%)
versus the geometry-averaged nuclear modification \rdauave, there are constrained parametric dependencies for the linear,
exponential, and quadratic cases.  In~\cite{Adare:2010fn}, the
analytic parameterization for \mrt as a function of the average \lrt was used to compare
the nuclear modification in all centralities for a given parameter $a$ value.  
For a given geometric dependence, varying the values of $a$ results in a locus 
of points for a constrained relationship between \rdau and \rcp.  Shown
in Figure~\ref{fig_expcase} as dotted, solid, and dashed black lines are the
result of that analytic calculation for the exponential, linear, and quadratic
cases respectively.  
To be clear, these curves are calculated purely from a Monte Carlo
Glauber geometry, the average density-weighted nuclear thickness \lrt,
and the simple geometric dependence equation (i.e. no specific model of nPDFs, \sbr, etc).
Also shown are the PHENIX experimental data with the lines
as point-to-point uncorrelated uncertainties and the ellipses as one-standard
deviation contours from the combined systematic uncertainties.  As stated in~\cite{Adare:2010fn},
this demonstrates that the forward rapidity \jpsi data cannot be reconciled with
an exponential or linear geometric dependence for the nuclear modification.

We test this picture further by additionally plotting the results from our calculations using 
the EPS09 nPDFs and $\sigma_{br}$ model.
In Figure~\ref{fig_expcase}, we show all EPS09 nPDF parameter sets using the
exponential geometric dependence, a range of values of \sbr (from 0-18 mb in 2 mb steps) and the full 
 range of rapidity values (as red lines). The subset corresponding to $\sbr = 0$ are shown
as blue lines. As expected, since the nPDF dependence is
exponential, the blue lines fall almost perfectly on the analytic pure
exponential case (dotted black line). With the 
additional \sbr contribution which also has an exponential geometric dependence,
we expected that everything would collapse onto the same line. However, with two
competing effects \rdauave=1 does not always equate with the trivial
case of no modification, but can also occur if the two effects average
to 1.  In the latter case, \rcp need not be 1.  Specifically in our
case, in the
backward rapidity region the nPDF leads to an enhancement
(anti-shadowing) and the \sbr to a suppression. This competition can
lead to the case where \rdauave = 1, while the \rcp $\ne 1$
($ie.$ some slight enhancement in peripheral events due to the
nPDF effect and some slight suppression in central events due to the
\sbr effect).  This effect leads to the slight splitting of the lines for values
near \rdau = 1.

Shown in Figure~\ref{fig_lincase} are the same quantities for the nPDF
linear case in our calculation. Again, the case with $\sbr =
0$ leaves only the purely linear nPDF and thus the blue lines collapse
onto the analytic linear case (solid black curve). The red curves for all $\sbr >
0$ cases result in a geometric dependence that is part linear and part
exponential. Thus, one sees that for larger suppressions (also larger
\sbr values), the red curves move between the analytic linear case to
the analytic exponential case. One again sees some cases of \rdauave = 1, 
while the \rcp is not equal to one for the same
reason as described above.

Lastly, in Figure~\ref{fig_quadcase}, we show the quadratic case. In
this case with $\sbr = 0$ the blue lines are close to the black dashed
analytic quadratic case, but not perfectly. This is due to the
inclusion of fluctuations in the thickness included in our full
calculation shown here. As one increases the value of \sbr in 2 mb
increments one sees the red lines moving to the left as the
exponential geometric dependence from \sbr dominates over the
quadratic nPDF effect.

This full suite of curves reveals
that even attempting to fit just the forward rapidity data with a
larger and larger \sbr will not successfully capture the full
centrality dependence (even using the quadratic nPDF contribution).

The results of fitting \rcp versus rapidity shown in Figures~\ref{fig_rcp_data_line} and~\ref{fig_rcp_data_quad}
demonstrated that no variation in the
model (e.g. EPS09 nPDF parameter sets, nPDF geometric dependence, single \sbr values,
etc.) can be reconciled with the full rapidity and centrality dependence of the
experimental data.  It is possible that \sbr may have a rapidity dependence
due to the different relative velocity of the
\ccbar pair with respect to the target nucleons. A na\"{i}ve
expectation is that a shorter time spent in the nucleus should result
in a smaller \sbr due to a smaller growth in the physical size of the
\ccbar pair towards that of the final state \jpsi. However,
in order to attempt to reconcile the data with a rapidity-dependent \sbr
the cross section would need to be much larger at larger rapidity (as shown
in~\cite{Brambilla:2010cs} in Section 5.2).  In fact, the results from the
\rcp and \rdau correlation indicate that no value of \sbr (even with an
independent value for forward rapidity) can properly describe the centrality
dependence.  This fact can be understood by noting that no matter how large a value of \sbr one
uses, the geometric dependence is always exponential.

\begin{figure*}[!t]
  \includegraphics[width=0.45\linewidth]{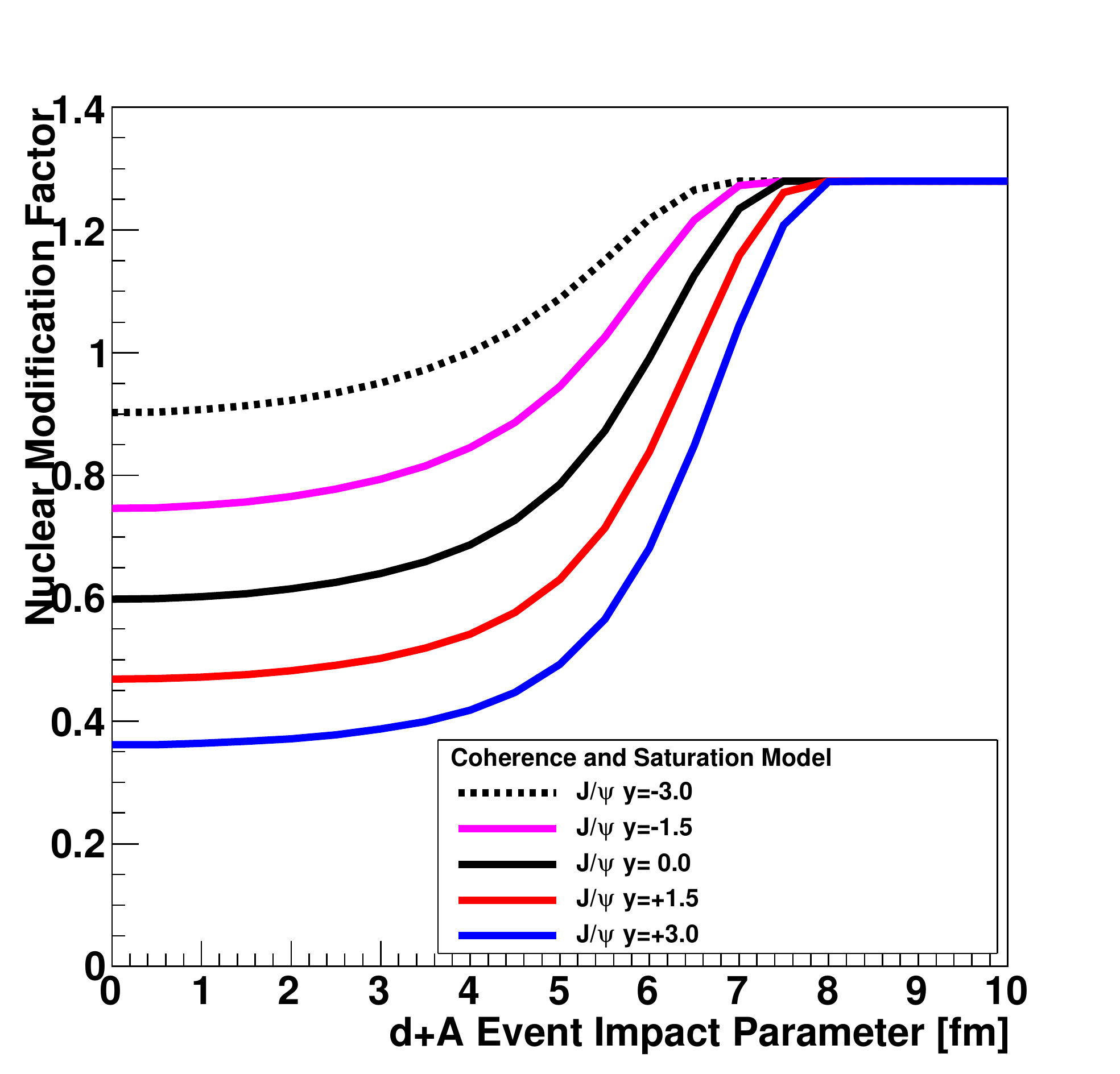} 
  \includegraphics[width=0.45\linewidth]{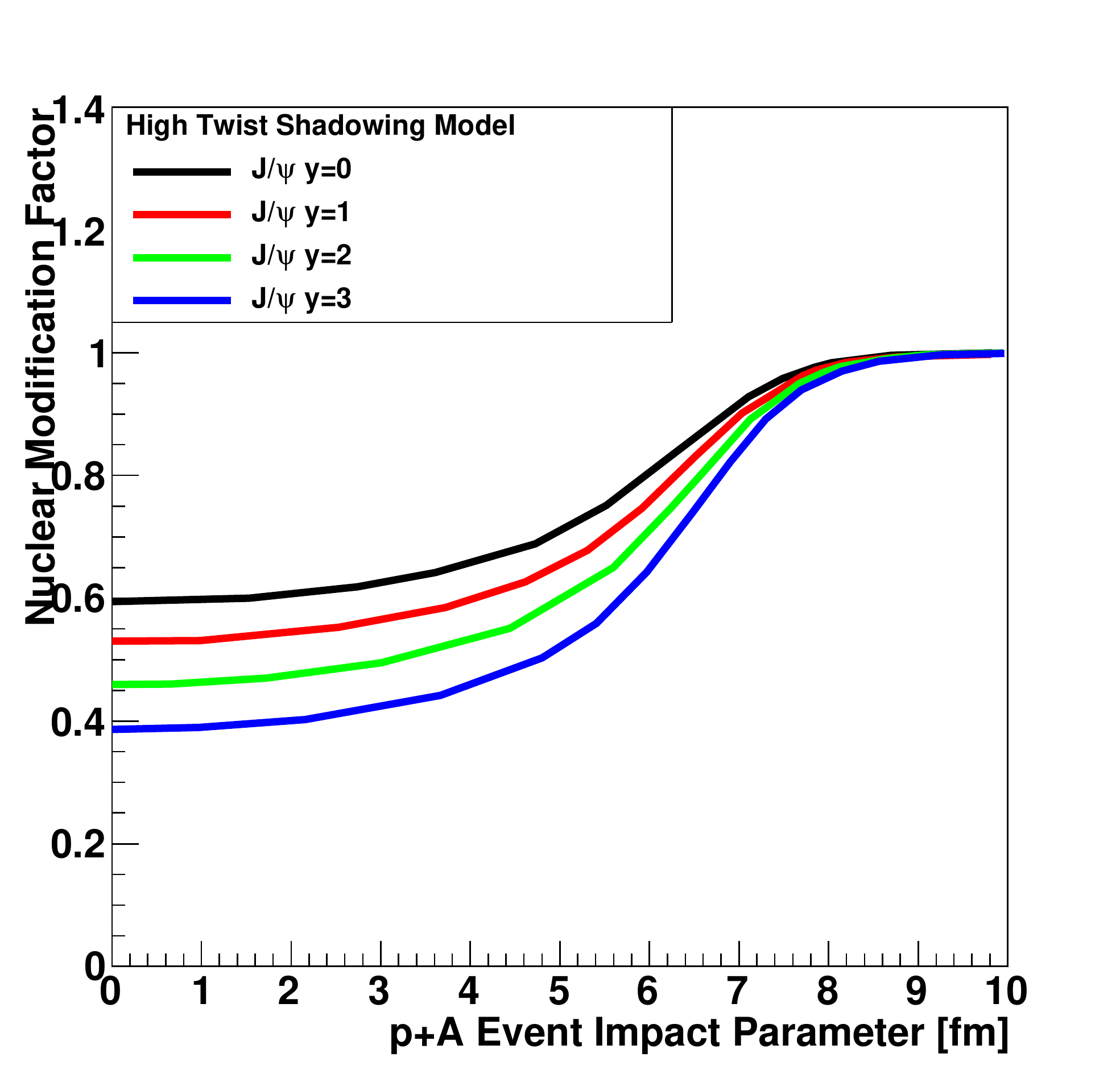} 
  \caption{
    (Left Panel) Shown are \jpsi nuclear modification results from the
    color glass condensate calculation including gluon saturation from~\cite{tuchin_private}
    as a function of event impact parameter in \dau reactions.
    (Right Panel) Shown are the results of a coherence and color transparency
    calculation~\cite{Kopeliovich:2010nw} for the \jpsi
    nuclear modification as a function of event impact parameter in \pau reactions.
  }
  \label{fig_coherence1}
\end{figure*}

Another possibility is that the geometric dependence
of the nPDF varies with rapidity. While this is possible, without
direct theoretical guidance or other experimental constraints, one would just 
be arbitrarily fitting the data.  In the next section we consider two additional
models in the literature that have different physics than this framework.

\section{Additional Model Comparisons}

One proposal for physics beyond the factorized nPDF and $\sigma_{br}$
involves calculations incorporating gluon saturation (non-linear evolution of the gluon
distributions at low $x$).  In the PHENIX paper~\cite{Adare:2010fn}, the data were
compared with a color glass condensate calculation~\cite{Kharzeev:2005zr} 
that incorporated suppression at low $x$ from gluon
saturation and enhancement from double-gluon exchange diagrams.
More recent calculations following this
framework~\cite{tuchin_private} include a more accurate treatment of the
nuclear geometry and the dipole-nucleus scattering amplitudes, and are
consistent with recent calculations for the Au+Au
case~\cite{Tuchin:2009tz}.  Shown in Figure~\ref{fig_coherence1} (left panel) is the calculated \jpsi nuclear
modification for different rapidities as a function of the \dau event
impact parameter ($b$).  It is notable that for large $b$ there is a 30\%
nuclear enhancement, even though the coherence is only an effect in the
longitudinal direction and the local nuclear density for these large-$b$
events is small.  

In an alternative calculation presented in~\cite{Kopeliovich:2010nw},
the \jpsi production is controlled by coherence and color transparency
effects.  Shown in Figure~\ref{fig_coherence1} (right panel) is the
calculated \jpsi nuclear modification factor as a function of event
impact parameter in \pau reactions.

\begin{figure}[htb]
  \includegraphics[width=0.9\linewidth]{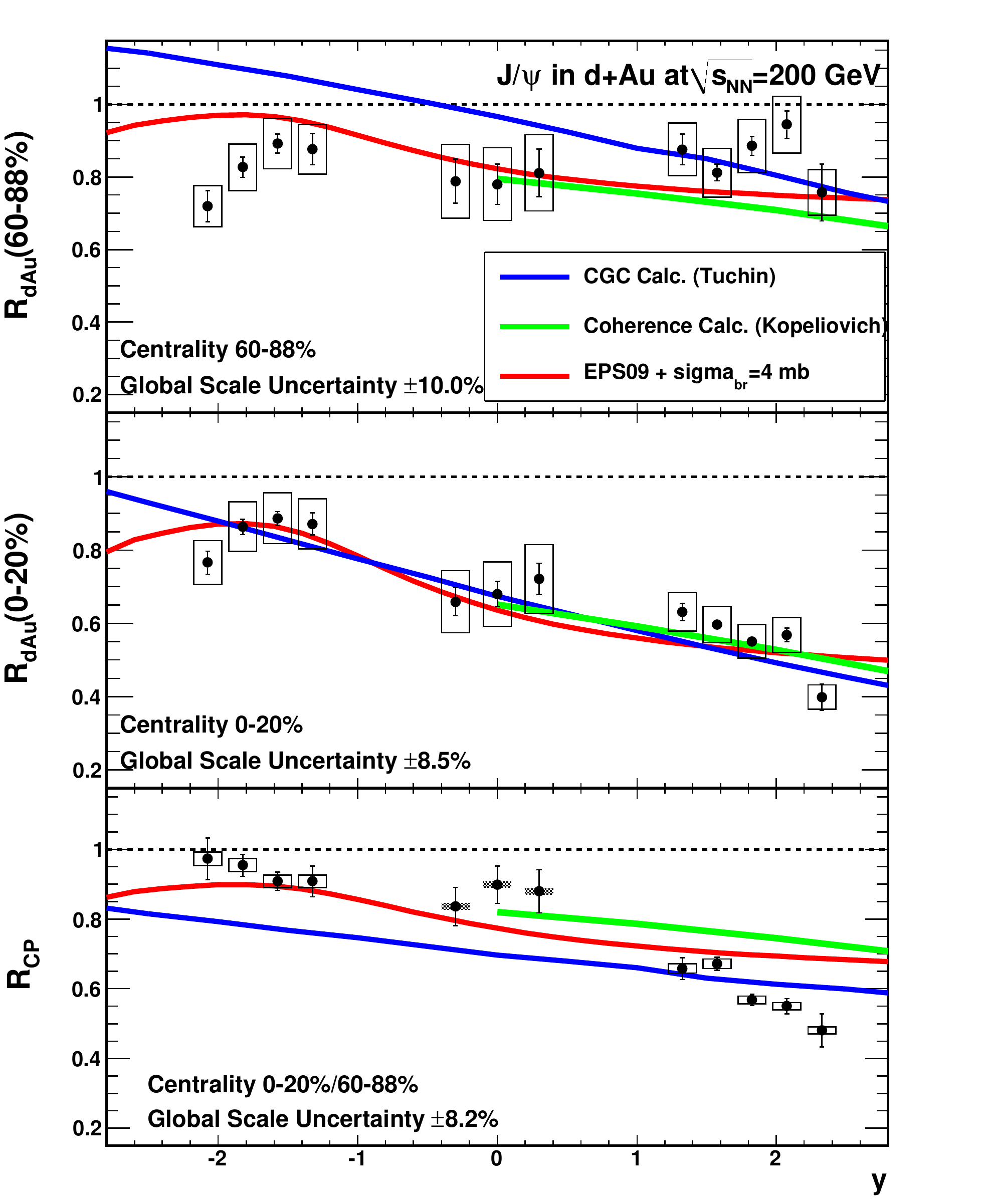} 
  \caption{
    Shown are the PHENIX experimental data for \jpsi \rdau peripheral (top) \rdau central (middle),
    and \rcp 0-20\%/60-88\% (bottom) as a function of rapidity.
    The green curve results from~\cite{Kopeliovich:2010nw} with coherence effects
    and color transparency. The blue curve is the
    color glass condensate calculation~\cite{tuchin_private}. The red curve 
    is our calculation with nPDF EPS09 default set=1 with a
    linear geometric dependence and $\sbr= 4$ mb.}
  \label{fig_coherence2}
\end{figure}

In either case, we can fold these dependencies with the \dau event impact
parameter distributions or the \rt distribution (for the \pau predictions), and compute
the \rdau and \rcp modification factors to compare with the PHENIX experimental data.  These comparisons
are shown in Figure~\ref{fig_coherence2}.  We also show for comparison the EPS09 nPDF default with linear 
geometric dependence and \sbr = 4 mb.  The calculation
from~\cite{Kopeliovich:2010nw} yields similar results to the EPS09 nPDF and \sbr calculation,
and has insufficient suppression at the most forward rapidity.  The color glass condensate calculation
shows better agreement at forward rapidity, though the rapidity dependence is not as steep as that of the \rcp 
experimental data. But it has  
substantial disagreement with the data at mid and backward
rapidities.  Note that that calculation assumes coherence over the entire longitudinal extent of the nucleus,
but this coherence approximation is no longer valid at some higher-$x$ values ($ie.$ moving towards
mid and then backward rapidity).  It is also no longer valid for low densities that occur at large impact
parameter values.  Thus, the result that 
for large impact parameters the color glass condensate  calculation approaches
$\rdau \approx 1.3$ (shown in the left panel of Figure~\ref{fig_coherence1}) is likely
to simply be outside the range of validity of the calculation.

\section{Initial-State Energy Loss}

Another physical effect that has been proposed is that of initial
state parton energy loss.  The very forward rapidity \jpsi are
produced from a high $x_{1}$ parton (from the deuteron) and a low
$x_{2}$ parton (from the gold nucleus).  If the high $x_{1}$ parton
from the deuteron loses some energy before the hard scattering, this
will result in a lower \jpsi production probability and a shift
backward in rapidity for any produced particles (including the \jpsi).
This framework has been used in an attempt to reconcile lower energy
Drell-Yan data in \pA collisions (see for example~\cite{Moss:2001hq,Johnson:2001xfa}).  
However, the same data have
also been interpreted in terms of nuclear shadowing models instead of
large initial-state energy loss.  It is unclear if these experimental
data sets can be reconciled with the same energy loss, and whether that
provides a consistent picture for all such $p(d)$+$A$ data.

More recently, in~\cite{Neufeld:2010dz} a calculation is presented of initial-state parton energy loss
and its impact on Drell-Yan production with predictions for measurements in $p+A$ collisions.  In the case
of initial-state radiative energy loss, they predict that $\Delta E/E \propto L$, where $L$ is the path through
the nucleus prior to the hard scattering.  Data from experiment E906 at Fermilab will directly address
this prediction, and in the clean Drell-Yan channel without final state effects~\cite{Reimer:2007iy}.

As a preliminary investigation of the impact of initial-state parton
energy loss, we have implemented this mechanism in our calculation (in
addition to the nPDF and \sbr contributions).  Within the Monte Carlo
Glauber, we also calculate $N_{tube}[before]$, the number of Au
nucleus nucleons in the tube that have a $z$ location prior to the $z$
position of the binary collision of interest.  We posit that the
initial-state energy loss is proportional to $N_{tube}[before]$, which
is the same as being proportional to the path $L$, with the inclusion
of local fluctuations.  For this calculation, we have utilized the
PYTHIA production $g+g \rightarrow \jpsi + X$ kinematics.  For each
binary collision location, we randomly select a PYTHIA $x_{1}$,
$x_{2}$ combination and the average \jpsi rapidity for those
kinematics.  We then calculate the expected $x_{1}$ shift due to the
energy loss corresponding to the particular $N_{tube}[before]$ value.
From this information, we calculate the decrease in the probability
for these partons to produce a \ccbar pair and the new (lower) average
final state \jpsi rapidity with the modified parton kinematics.  We
have varied the proportionality constant for the initial-state energy
loss in the $N_{tube}[before]$ dependence.

\begin{figure*}[htb]
  \includegraphics[width=0.45\linewidth]{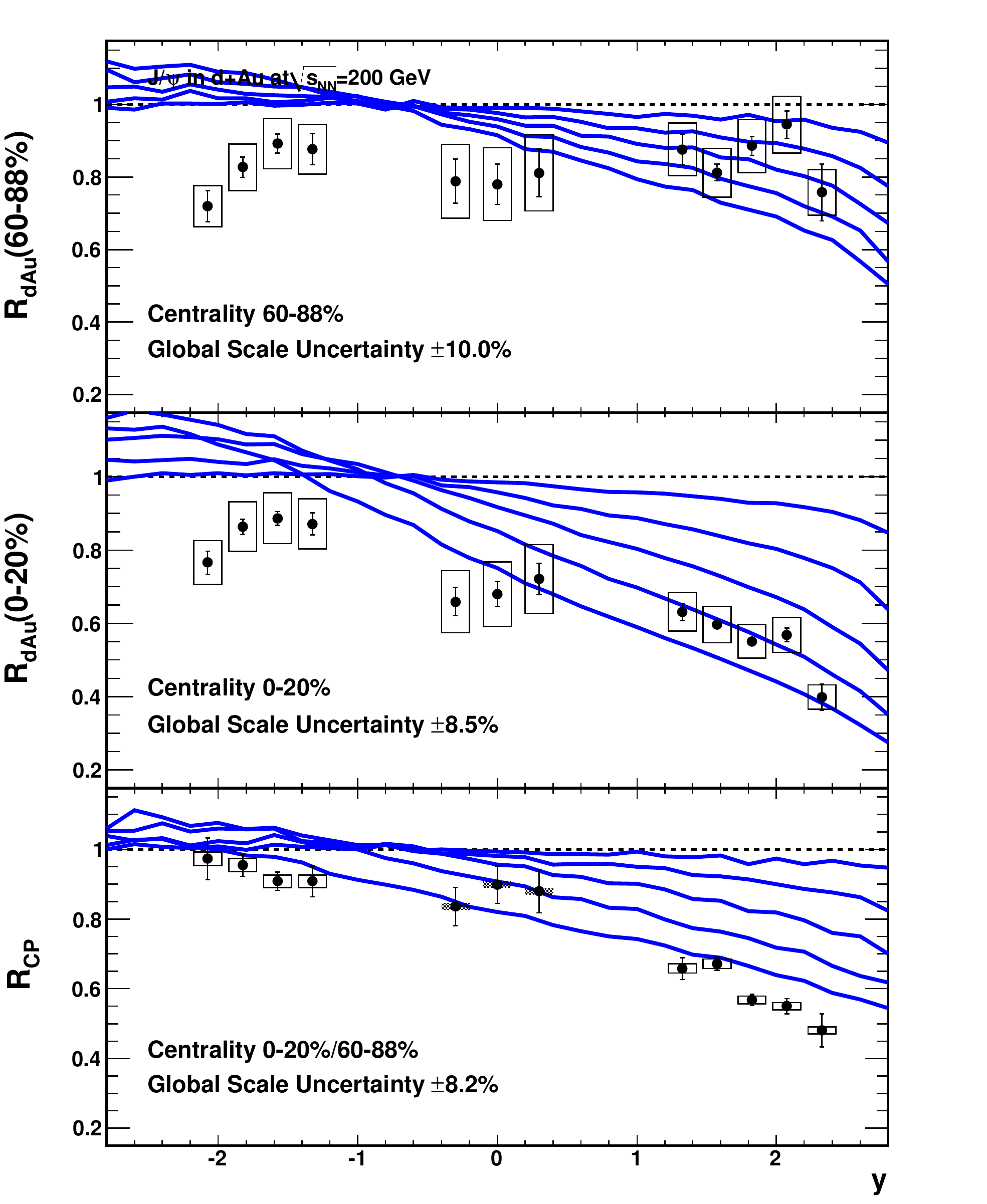}
  \includegraphics[width=0.45\linewidth]{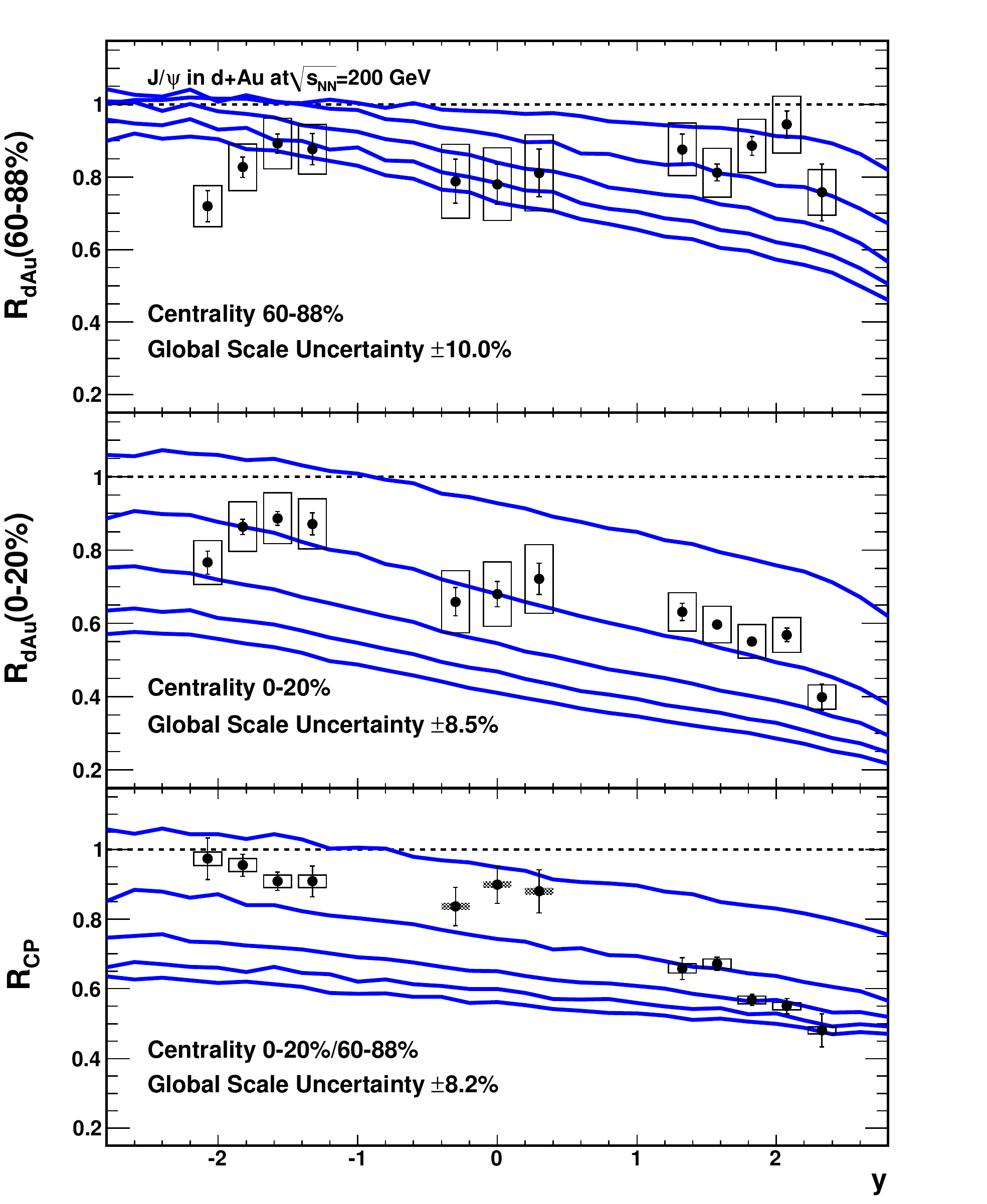} 
  \caption{
    (Left Panel) Calculation including initial-state parton energy loss only and with
$\Delta E/E \propto L$ or equivalently in our case with fluctuations to $N_{tube}[before]$.
The curves correspond to coefficients of 0.01/fm, 0.03/fm, 0.05/fm, 0.07/fm, and 0.09/fm (from
upper to lower in order).
    (Right Panel) Calculation including initial-state parton energy loss only and with
$\Delta E/E \propto L^{2}$ or equivalently in our case with fluctuations to $N_{tube}[before]^{2}$.
The curves correspond to coefficients of 0.005/fm$^{2}$, 0.015/fm$^{2}$, 0.025/fm$^{2}$, 0.035/fm$^{2}$, and 0.045/fm$^{2}$
(from upper to lower in order).
  }
  \label{fig_eloss_sample}
\end{figure*}

In Figure~\ref{fig_eloss_sample} (left panel), we show the calculation results including only initial
state parton energy loss (i.e. no nPDF modification and \sbr = 0).  One observes a larger suppression
at foward rapidity, and in fact a modest enhancement at backward rapidity.  In Figure~\ref{fig_eloss_sample} (right panel),
we show the results when now assuming a quadratic path dependence for the energy loss (i.e.  $\Delta E/E \propto L^{2}$).
In this case, for large coefficients in the proportionality constant, there is large suppression at all rapidities.
For either the $L$ or $L^{2}$ dependence, one cannot achieve good agreement with the experimental data with 
initial-state parton energy loss alone.

\begin{figure*}[htb]
  \includegraphics[width=0.45\linewidth]{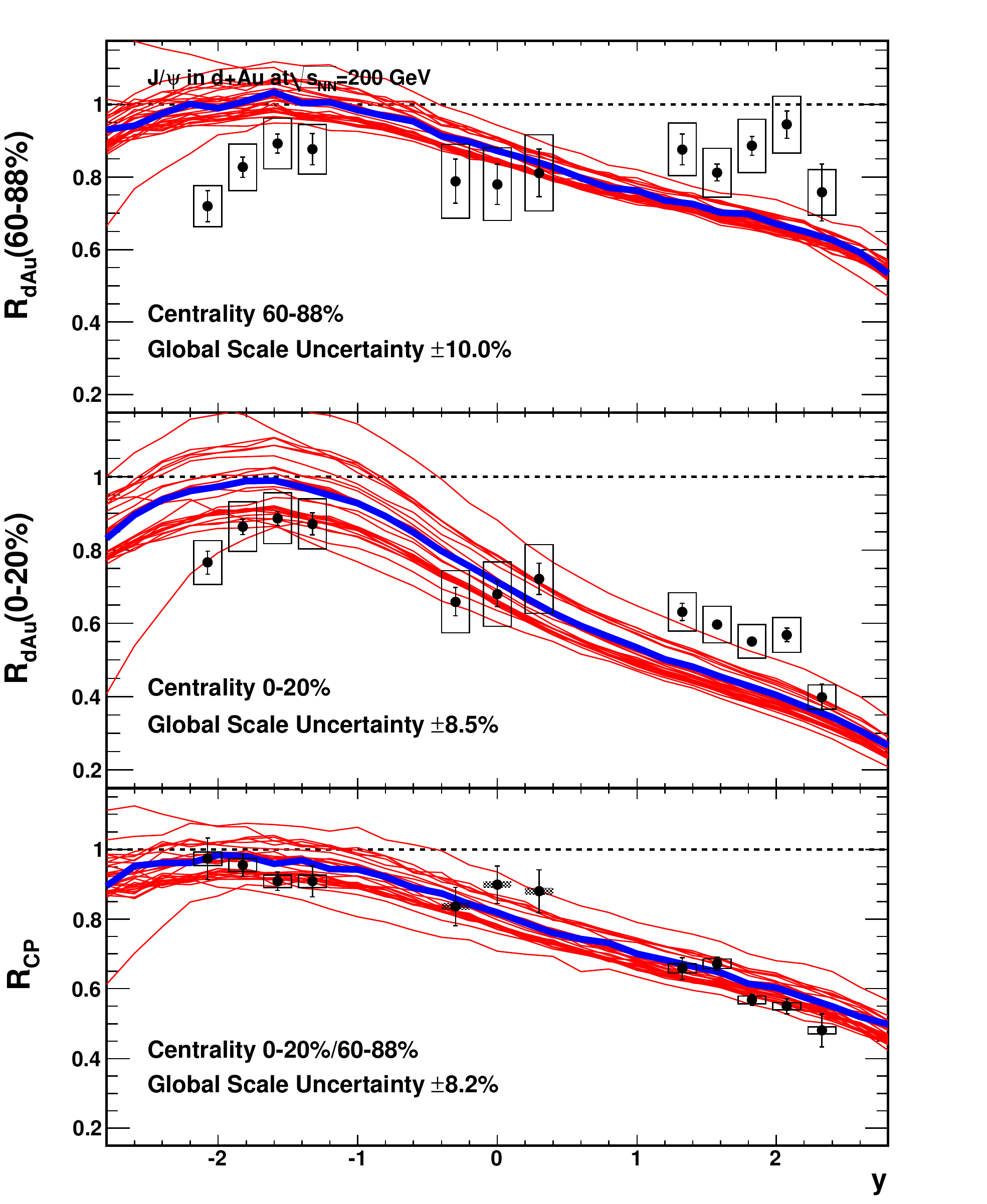} 
  \includegraphics[width=0.45\linewidth]{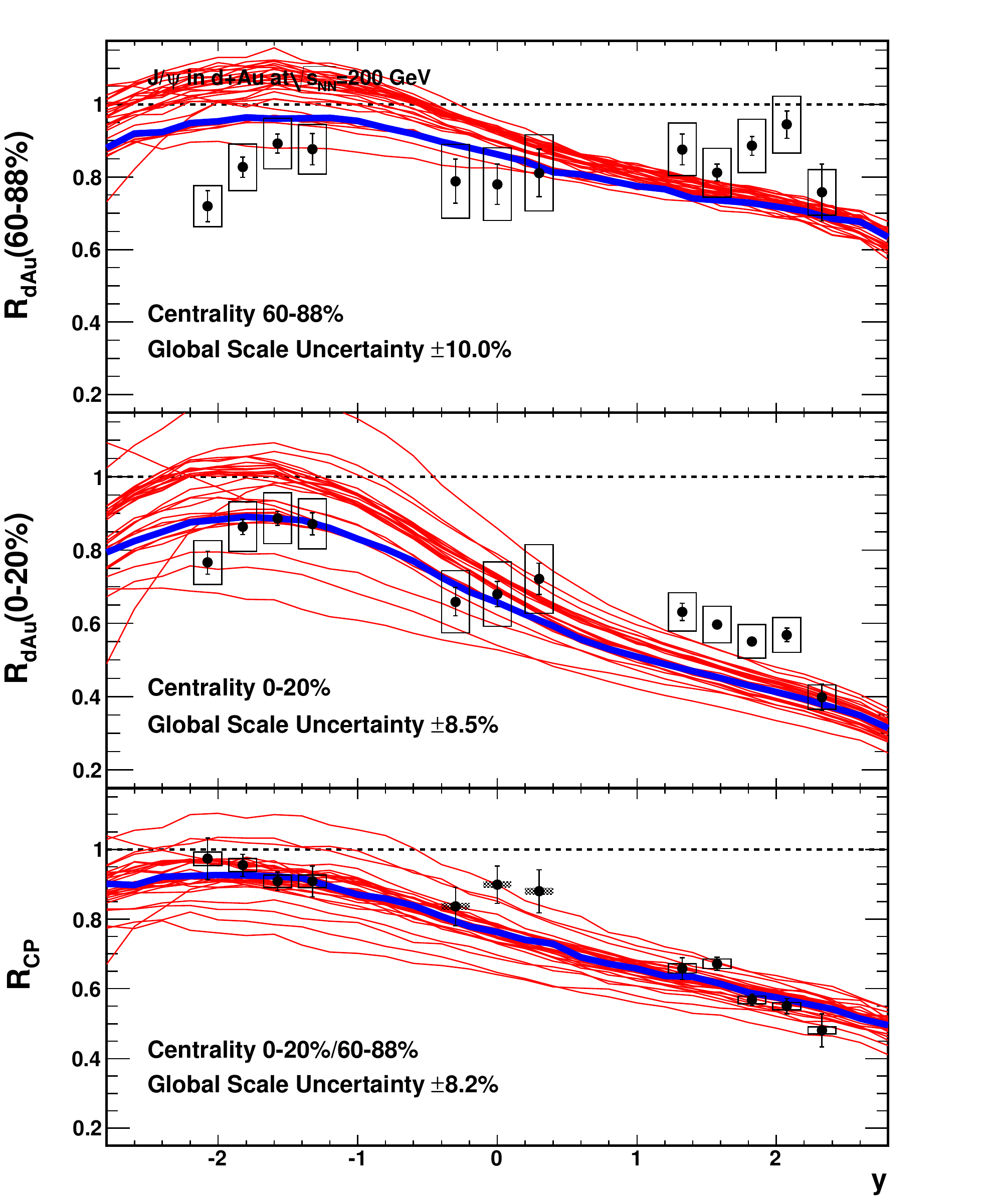} 
  \caption{
    (Left Panel) Best fit including initial-state parton energy loss assuming
the loss is proportional to $N_{tube}[before]$.
    (Right Panel) Best fit including initial-state parton energy loss assuming
the loss is proportional to $N_{tube}[before]^{2}$.
  }
  \label{fig_eloss}
\end{figure*}

We then include all three nuclear effects (nPDF modification, \sbr, and initial-state parton energy loss), and
vary the EPS09 nPDF parameter set and fit for the best \sbr and energy loss coefficient.
In Figure~\ref{fig_eloss} (left panel), we have found the best $\tilde{\chi}^{2}$ from the fit to 
the \rcp.  The blue curve represents
the best overall fit for all EPS09 parameterizations and the optimal \sbr and initial-state energy loss
coefficient.  The best fit gives a reasonable description of the \rcp and corresponds to EPS09 parameter set 23, \sbr = 3 mb,
and $\Delta E/E \approx 0.05/$fm$ \times L$ (converting the average $N_{tube}[before]$ to a length
through normal nuclear matter density).  The $\tilde{\chi}^{2}$ = 20.2 which is a better fit than without
the initial-state energy loss, but still gives a probability of less than 5\%.  However, more striking is that
with a reasonable fit to the \rcp, there is no global agreement with the overall suppression for \rdau in peripheral
or central events.

In Figure~\ref{fig_eloss} (right panel), we show the same quantities, but now assuming that
the initial-state energy loss is quadratic in the path or $N_{tube}[before]$.  In this case the best
fit to \rcp has EPS09 parameter set 5, \sbr = 4 mb, and the initial-state energy loss corresponds to approximately
$\Delta E/E \approx 0.005/$fm$^{2} \times L^{2}$.  Again there is no overall global good fit for \rdau and \rcp.  

These results are a first look in comparing the simplest initial-state energy loss calculation to these
\jpsi data.  We have not included Poisson fluctuations of the radiated quanta, which are important
when one is near the very high-$x_{1}$ limit as pointed out in~\cite{Neufeld:2010dz}.  However, in their
calculation they have not included the important fluctuations in the $L$ value itself, as we have done
through utlizing $N_{tube}[before]$.  It is premature to draw any firm conclusions about the implications
for initial-state parton energy loss from these comparisons with \jpsi data alone.  Further
calculations are necessary, as well as comparisons to observables from other collision energies and other 
final state produced particles.

\section{Summary}

In this paper, we have presented an extension of calculations for \jpsi
nuclear modifications including modified parton distribution functions (nPDFs)
and fit parameter \sbr.  Utilizing the full set of EPS09 nPDFs and three
different postulated geometric dependencies, we find that the calculations
cannot be reconciled with the full rapidity and centrality dependence of the
PHENIX \dau \ \jpsi data for any nPDF variation and any \sbr value.  Additionally,
the comparison with \rcp versus \rdau indicates that even a much larger \sbr
at forward rapidity cannot reconcile the calculation with the data, since the
\sbr contribution always has an exponential geometric dependence.  We also compare
to two different coherence calculations and find no agreement across all rapidities
and centralities.  Most likely new physics beyond these calculations is a signficant
contributor, perhaps initial-state parton energy loss.  A first look at a simple
parameterization of initial-state energy loss allows a better description of the \jpsi
\rcp, but without a good simultaneous description of \rdau.  Additional constraints
from Drell-Yan and direct photon observables at forward rapidity and at different $\sqrt{s_{NN}}$ energies
may be necessary to disentangle these effects.

\begin{acknowledgments}
  We thank Kirill Tuchin for providing us with the color glass
  condensate calculation results and useful discussions, and Michael
  Stone for generating the PYTHIA event samples.  We also acknowledge
  useful discussions with Mike Leitch, Darren McGlinchy, and Ramona
  Vogt.  JLN and MGW acknowledge funding from the Division of Nuclear
  Physics of the U.S. Department of Energy under Grant
  No. DE-FG02-00ER41152.  LALL acknowledges that this work was
  performed under the auspices of the U.S. Department of Energy by
  Lawrence Livermore National Laboratory under Contract
  DE-AC52-07NA27344. ADF acknowledges funding from the National Science 
  Foundation under contract PHY-07-56474. We also thank the Institute for Nuclear Theory
  at the University of Washington for its hospitality and the
  Department of Energy for partial support during some of this work.
\end{acknowledgments}

\bibliographystyle{apsrev}   
\bibliography{geom_suppression}   

\end{document}